%% file: masters_paper.tex
\documentclass[12pt]{article} 
\usepackage{fullpage}
\usepackage{amssymb}
\usepackage{amsthm}
\usepackage{amsmath} %\usepackage{amstext}
\usepackage[latin1]{inputenc} 
\usepackage{hyperref} % NOTE: this may cause you to have to hit compile a couple times initially.
\usepackage{xargs}
\usepackage{xifthen}

\usepackage{algorithmic}

\usepackage{times} 
\usepackage{units}

\usepackage{graphicx}

\newenvironment{ppe}{ \begin{enumerate}
  \vspace{-5pt}
  \setlength{\itemsep}{1pt} 
  \setlength{\parskip}{0pt}
  \setlength{\parsep}{0pt}
}{\end{enumerate}}
\newenvironment{ppi}{ \begin{itemize}
  \vspace{-5pt}
  \setlength{\itemsep}{1pt} 
  \setlength{\parskip}{0pt}
  \setlength{\parsep}{0pt}
}{\end{itemize}}

\newenvironment{packed_item}{ \begin{itemize}
  \setlength{\itemsep}{1pt} 
  \setlength{\parskip}{0pt}
  \setlength{\parsep}{0pt}
}{\end{itemize}}

\newtheorem{theorem}{Theorem} 
\newtheorem{cor}{Corollary} 
\newtheorem{prop}{Proposition}
\newtheorem{lemma}{Lemma}
\newtheorem{fact}{Fact}

\theoremstyle{definition}
\newtheorem{defn}{Definition}

\newenvironment{remark}[1][Remark]{\begin{trivlist} \item[\hskip \labelsep
{\bfseries #1}]}{\end{trivlist}}

\newcommand{\DEF}{{\downarrow}}
\newcommandx*{\ran}[2][1={},2={}]
 {
  \ifthenelse{ \isempty{#2} }
             { [#1] }  
             { [#1\,{:}\,#2] }
}

\newcommand{\Ufixed}{U_{{\sf fixed}}}

\newcommand{\UL}{U_{\mathsf{L}}}

\newcommand{\Vars}{\mathsf{Vars}}

\newcommand{\node}{{\sf node}}
\newcommand{\var}{{\sf var}}
\newcommand{\Par}{{\sf par}}

\newcommand{\adv}{{\sf adv}}

\newcommand{\sibl}{{\sf sibl}}

\newcommand{\Learnable}{{\sf Learnable}}
\newcommand{\RightThrifty}{{\sf RightThrifty}}
\newcommand{\LeftThrifty}{{\sf LeftThrifty}}

\newcommandx*{\rPath}[2][1=I,2=r]
 {
  \textsf{Path}^{#1}(#2)
 }
\newcommand{\BnPath}{{\sf BnPath}}
\newcommand{\SiblBnPath}{{\sf SiblBnPath}}
\newcommand{\RightPath}{{\sf RightPath}}

\def\advk{{\sf adv}_k}
\def\advpi{{\sf adv}_{\pi}}

\newcommand{\isc}{i_{{\sf sc}}}
\newcommand{\ginit}{g_{{\sf init}}}
\newcommand{\Ginit}[1]{G_{{\sf init}}^{#1}}
\def\IntAdv{\text{{\sc InterAdv}}}
\def\Fill{\text{{\sc Fill}}}
\def\LearnNode{\text{{\sc LearnNode}}}

\renewcommand{\algorithmiccomment}[1]{//\ \emph{#1}}
\renewcommand{\algorithmicrequire}{\textbf{Precondition:}} 
\renewcommand{\algorithmicensure}{\textbf{Postcondition:}}

\def\s{\vva}
\def\gets{\leftarrow}

\def\spi{\vva_{\pi}}
\def\sk{\vva_{k}}

\newcommand{\vva}{ \vec{a} }
\newcommand{\vvb}{ \vec{b} }

\newcommandx*{\FT}[2][1=h,2=k]{FT^{#1}(#2)}
\newcommandx*{\BT}[2][1=h,2=k]{BT^{#1}(#2)}
\newcommand{\Bpebbles}{\ensuremath{\mathsf{\#Bpebbles}}}
\newcommand{\BWpebbles}{\ensuremath{\mathsf{\#BWpebbles}}}
\newcommand{\FRpebbles}{\ensuremath{\mathsf{\#FRpebbles}}}

\newcommandx*{\dFstate}[2][1=h,2=k]{\ensuremath{\mathsf{\#detFstates}^{#1}({#2})}}
\newcommandx*{\dBstate}[2][1=h,2=k]{\ensuremath{\mathsf{\#detBstates}^{#1}({#2})}}
\newcommandx*{\nFstate}[2][1=h,2=k]{\ensuremath{\mathsf{\#ndetFstates}^{#1}({#2})}}
\newcommandx*{\nBstate}[2][1=h,2=k]{\ensuremath{\mathsf{\#ndetBstates}^{#1}({#2})}}

\newcommand{\Dom}{{\sf Dom}}

\newcommand{\nl}{\ensuremath{\mathbf{NL}}}
\newcommand{\logspace}{\ensuremath{\mathbf{L}}}

\newcommand{\logdcfl}{\ensuremath{\mathbf{LogDCFL}}}
\newcommand{\np}{\ensuremath{\mathbf{NP}}}

\newcommand{\lb}{\left <}
\newcommand{\rb}{\right >}
\newcommand{\pair}[1]{\lb #1 \rb}
\newcommand{\ksq}{[k]{\times}[k]}

\title{Pebbling and Branching Programs Solving the Tree Evaluation Problem}
\author{Dustin Wehr}
\begin{document}
\maketitle

\begin{abstract}
We study restricted computation models related to the \emph{tree evaluation problem}. 
The TEP was introduced in earlier work as a simple candidate for the (\emph{very}) long term
goal of separating $\mathbf{L}$ and $\mathbf{LogDCFL}$.
The input to the problem is a rooted, balanced binary tree of height $h$,
whose internal nodes are labeled with binary functions
on $[k]=\{1,\ldots,k\}$ (each given simply as a list of $k^2$ elements of $[k]$), 
and whose leaves are labeled with elements
of $[k]$. Each node obtains a value in $[k]$ equal to its binary 
function applied to the values of its children.  The output is the
value of the root. 
The first restricted computation model, called \emph{fractional pebbling}, is a generalization
of the black/white pebbling game on graphs, and arises in a natural way from the search for 
good upper bounds on the size of nondeterministic branching programs solving the TEP - for 
any fixed $h$, if the binary tree of height $h$ has fractional pebbling cost at most $p$, then there are 
nondeterministic branching programs of size $O(k^p)$ solving the height $h$ TEP.
We prove a lower bound on the fractional pebbling cost of $d$-ary trees that is
tight to within an additive constant for each fixed $d$. 
The second restricted computation model we study is a semantic restriction on (non)deterministic 
branching programs solving the TEP -- \emph{thrifty} branching programs. Deterministic (resp. nondeterministic) thrifty BPs suffice to implement the best known algorithms, based on black pebbling (resp. fractional pebbling), for the TEP. In earlier work, for each fixed $h$ a lower bound on the size 
of thrifty deterministic branching programs was proved that is tight for sufficiently large $k$. We give an alternative proof that achieves the same bound for all $k$ and $h$. We show the same bound still holds in a less-restricted model, and also that gradually weaker lower bounds can be obtained for gradually weaker restrictions on the model.
\end{abstract}

\section{Introduction}

The motivations for this paper are those of \cite{fsttcs}, and the goals are to extend and improve on the results given there (with the exception of Theorem \ref{t:fractLB}, which appeared there verbatim). But from a wider view, what we want is to improve our understanding of $\logspace$ in the hope that this will help in eventually separating it from (apparently) larger classes. We study the tree evaluation problem (TEP), which was defined in \cite{mfcs} and shown to be in $\logdcfl$.

The function version of the \emph{Tree Evaluation problem} $\FT$ is 
defined as follows. Let $T^h$ be the balanced binary tree of height $h$ (see Fig. \ref{sample}). 
For each internal node $i$ of $T^h$ the input includes a function
$f_i: \ksq \to [k]$ specified as $k^2$ integers in
$[k]=\{1,\ldots,k\}$. For each leaf the input includes an integer in $[k]$.
We can then say that each internal tree node takes a value in $[k]$ by applying its
function to the values of its children. The function
problem $\FT$ is to compute the value of the root, and the decision version 
$\BT$\ is to determine whether this value is $1$.

Since $\BT \in \logdcfl$, it is not hard to show that for {\em any} unbounded function $r(h)$, 
a lower bound of $\Omega(k^{r(h)})$ on the number of states for deterministic
(resp. non-deterministic) branching programs solving $\FT$ or $\BT$ would 
separate \logdcfl\ and \logspace\ (resp. \nl) \footnote{Of course, doing so would
actually yield the stronger result: Nonuniform $\logspace \not \subseteq \logdcfl$ (resp. 
Nonuniform $\nl \not \subseteq \logdcfl$).}. To see this, note that inputs to 
$\BT$ can be encoded with $(2^{h-1}-1)k^2 \log k + 2^{h-1} \log k + O(1) = O(2^h k^2 \log k)$ 
bits, so it suffices to consider polynomial bounding function that are the product of a polynomial 
in $2^h$ and a polynomial in $k$, which $k^{r(h)}$ is not.

In \cite{mfcs}, the TEP was defined more-generally on balanced $d$-ary trees, 
where the functions attached to internal nodes are of type $[k]^d \to [k]$. 
The motivation was that tight lower bounds for height 3 and all $d$ can be proved \cite{mfcs}, 
and proving the conjectured lower bound of $\Omega(k^7 / \log k)$ 
states (with $h = 4$ and $d = 3$ fixed, so that the input size $n(k)$ is $O(k^3 \log k)$ bits or $O(k^3)$\, $[k]$-valued variables) for unrestricted deterministic BPs 
would beat the best known lower bound of $\Omega( n^2 / (\log n)^2)$ states for a problem in $\np$, 
achieved using Ne\u{c}iporuk's method \cite{ne66}. Since we are focusing on restricted computation models
here, there is little to gain in including the parameter $d$. That being said, the fractional 
pebbling lower bound proved in Section \ref{s:fractLB} \emph{is} given for arbitrary $d$.

\bigskip

\section{Preliminaries}\label{s:preliminaries}

\input{preliminaries}

\section{Thrifty Branching Programs and Pebbling} \label{s:thrifty_branching_programs_and_pebbling}

\input{upperbound_and_height2_and_minDepth}

\input{thrifty_advice}

\section{Main Results}

\subsection{Fractional Pebbling Lower Bound}\label{s:fractLB}
\input{fractional_lower_bound}

\subsection{Less-Thrifty Branching Programs}\label{s:less-thrifty_branching_programs}
\input{less_thrifty}

\subsubsection{Less-Thrifty BPs with Additional Queried Variables}
\input{general_less_thrifty}

\section{Open Problems} \label{s:open_problems}

The first is a problem that can, in principle, be resolved using a computer. 

{\bf 1.} Show that for some $k,h$ there is a deterministic branching program with fewer than $(k+1)^h$ states that solves $\FT$. \\ \\ 
Theorem \ref{t:thrifty_advice} suggests the following conjecture: for all $h$, nondeterministic thrifty branching programs solving $\FT$ require $\Omega(k^{\FRpebbles(T^h)})$ states. 

{\bf 2.} Refute it, with or without the thrifty restriction. 

\bibliographystyle{alpha}
\bibliography{masters_paper}

\end{document}

%% file: preliminaries.tex
We write $[k]$ for $\{1,2,\ldots,k\}$.
For $h \ge 1$ we use $T^h$ to denote the balanced binary tree
of height $h$.

\medskip

\noindent
{\bf Warning:}  Here the {\em height} of a tree is the number
of levels in the tree, as opposed to the distance from root to
leaf.  Thus $T^2_2$ has just 3 nodes.

\medskip

\begin{figure}
\vspace*{.3cm}
\hspace*{5cm}\includegraphics[scale=0.40]{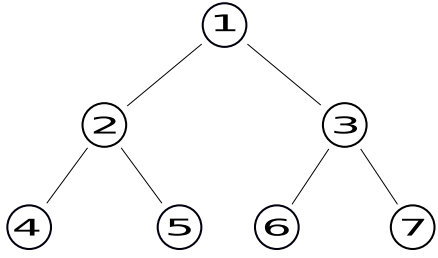}
\hspace*{1cm} \vspace*{.2cm}
\caption{The height 3 binary tree $T^3$ with nodes numbered heap style.}
\label{sample}
\end{figure}

\noindent
We number the nodes of $T^h$ as suggested by the heap data
structure.  Thus the root is node 1, and in general the children
of node $i$ are nodes $2i,2i+1$
(see Figure \ref{sample}). 

\begin{defn}[Tree evaluation problems]
\label{d:treeEval} \
\begin{quote}
An input $I$ for either the function or decision version of the problem includes: 
for each internal node $i$ of $T^h$, a function $f_i^I : \ksq \to [k]$ represented 
as $k^2$ integers in $[k]$, and for each leaf node $i$, an integer $l_i^I \in [k]$.

\emph{Function evaluation problem} $\FT$: On input $I$, compute
the value $v_1^I \in [k]$ of the root $1$ of $T^h$, where in general
$v_i^I = l_i^I$ if $i$ is a leaf and $v_i^I=f_i^I(v^I_{2i}, v^I_{2i+1})$ if $i$
is an internal node.

\emph{Boolean evaluation problem} $\BT$: Accept $I$ iff $v^I_1=1$.

\end{quote}
\end{defn}

\subsection{Branching programs}\label{s:branching_programs}

We use the same branching program model as in \cite{fsttcs} and \cite{mfcs}.

\begin{defn}[Branching programs]
  A \emph{nondeterministic $k$-way branching program} $B$ computing a
  total function $g:[k]^m\rightarrow R$, where $R$ is a finite set, is
  a directed rooted multi-graph whose nodes are called {\em states}. Every
  edge has a label from $[k]$.  Every state has a label from $[m]$,
  except $|R|$ {\em output} sink states consecutively labeled with the
  elements from $R$.  An input
  $(x_1,\ldots,x_m)\in [k]^m$ activates, for each $1\leq j\leq m$,
  every edge labeled $x_j$ out of every state labeled $j$. A
  {\em computation path} on input $\vec{x}=(x_1,\ldots,x_m)\in [k]^m$
  is a directed path consisting of edges activated by $\vec{x}$
  which begins with the
  unique start state and either ends in the final state labeled
  $g(x_1,\ldots,x_m)$ or is infinite.
  At least one such computation must end.
  The \emph{size} of $B$ is its number of states.  $B$ is
  \emph{deterministic $k$-way} if every non-output state has precisely
  $k$ outedges labeled $1,\ldots,k$. % \REMOVE{$B$ is \emph{binary} if $k=2$}

We say that $B$ solves a decision problem (relation) if it computes
the characteristic function of the relation.
\end{defn}

A $k$-way branching program computing $\FT$ or $\BT$ requires
$k^2$ $k$-valued arguments for each internal node $i$ of $T^h$ in
order to specify the function $f_i$, together with one $k$-valued
argument for each leaf. Thus in the notation of the above definition,
$\FT : [k]^m \rightarrow R$ where $R=[k]$ and
$m=(2^{h-1}-1) k^2 + 2^{h-1}$.
Also $\BT : [k]^m \rightarrow \{0,1\}$. 

{\bf Important:} Since we only study the tree evaluation problem (TEP) here, we give 
the input variables mnemonic names: $f_i(a,b)$ is an input variable 
(called an \emph{internal node variable}) for every internal 
node $i$ and $a,b \in [k]$ and $l_i$ is an input variable (called a \emph{leaf variable})
for every leaf $i$. 

For fixed $h$ we are interested in how the number of states
required for a $k$-way branching program to compute $\FT$ and $\BT$
grows with $k$. This is why we write $h$ in the superscript of $\FT$ 
and $\BT$. We define \dFstate\ (resp. \nFstate)
to be the mininum number of states required for a deterministic
(resp. nondeterministic) $k$-way branching program to solve $\FT$.
Similarly we define \dBstate\ and \nBstate\
to be the number of states required to $\BT$.

Thrifty programs are a restricted form of $k$-way branching
programs for solving tree evaluation problems, introduced in
\cite{fsttcs}.  Thrifty
programs efficiently simulate pebbling algorithms, and
implement the best known upper bounds for 
\nBstate\ and \dFstate, and are within a factor of $\log k$
of the best known for \dBstate.

\begin{defn}[Thrifty branching program]
\label{d:thrifty}
A deterministic $k$-way branching program which solves $\FT$
or $\BT$ is {\em thrifty} if during the computation on any input every
query $f_i(a,b)$ to an internal node $i$ of $T^h$ satisfies the
condition that
$\pair{a,b}$ is the tuple of correct values for the children of node $i$ 
(i.e. $v_{2i}^I =a$ and $v_{2i+1}^I = b$).
A non-deterministic such program is {\em thrifty} if for every input
every computation which ends in a final state
satisfies the above restriction on queries.
\end{defn}

This is a strong restriction. For example, a deterministic thrifty BP cannot, for 
any internal node $i$, iterate over all the $k^2$ variables that define $f_i$, or even 
just two distinct $f_i$ variables. 

In \cite{fsttcs} the following theorem is given, showing how upper bounds for
black pebbling and fractional pebbling yield upper bounds for deterministic and 
nondeterministic branching programs solving the TEP. The proof can be found in
\cite{manuscript}.

\smallskip

\noindent 
{\bf Theorem (\cite{fsttcs}): }
\begin{ppe}
 \item[(i)] If $T^h$ can be black pebbled with $p$ pebbles, then deterministic 
thrifty branching programs with $O(k^{p})$ states can solve $\FT$ and $\BT$.
 \item[(ii)] If $T^h$ can be fractionally pebbled with $p$
pebbles then non-deterministic thrifty branching programs can
solve $\BT$ with $O(k^p)$ states.
\end{ppe}

Also in \cite{fsttcs}, the following lower bound was given for deterministic thrifty programs. The proof can be found in \cite{manuscript}.
\smallskip

\noindent 
{\bf Theorem (\cite{fsttcs}): }  
For all $h$, for $k > {2^h \choose h-1}$ every deterministic thrifty branching program solving $\BT$ requires at least $\frac{1}{2} k^h$ states.

\smallskip

Theorem \ref{t:thrifty_advice} in Section \ref{s:thrifty_branching_programs_and_pebbling}, which is a special case of Theorem \ref{t:less_thrifty} in Section \ref{s:less-thrifty_branching_programs}, gives a small improvement on that result. The main improvement is that it gives a tight bound that holds for all pairs $k$ and $h$, rather than requiring that $k$ be much larger than $h$. The constant $1/2$ also goes away:
\smallskip 
\noindent 
{\bf Theorem \ref{t:thrifty_advice} :}
 For all $h,k$ every deterministic thrifty branching program solving $\BT$ requires at least $k^h$ states.

\subsection{Pebbling}\label{s:pebling}

The pebbling game for dags was defined by Paterson and Hewitt
\cite{pahe70} and was used as an abstraction for deterministic
Turing machine space in \cite{co74}.  Black-white pebbling
was introduced in \cite{cose76} as an abstraction of
non-deterministic Turing machine space (see \cite{nordstrom}
for a recent survey). Fractional pebbling was introduced in
\cite{fsttcs}.   

Let us first define three versions of the pebbling game. We will
not be proving anything about black-white pebbling directly, but 
fractional pebbling is a generalization of black-white pebbling, 
so it will be easier to define it first. 
The first is a simple `black pebbling' game:  A black
pebble can be placed on any leaf node, and in general
if all children of a node $i$ have pebbles, then one of the
pebbles on the children can be slid to $i$ (this is a
``black sliding move')'.  Any black pebble can be removed
at any time.  The goal is to pebble the root, using as few
pebbles as possible.  The second
version is `whole' black-white pebbling as defined in
\cite{cose76} with the restriction that we do not allow
``white sliding moves''.  Thus if node $i$ has a white
pebble and each child of $i$ has a pebble (either black or white)
then the white pebble can be removed.  (A white sliding move
would apply if one of the children had no pebble, and the
white pebble on $i$ was slid to the empty child.  We do not
allow this.)  A white pebble can be placed on any node at
any time.  The goal is to start and end with no pebbles,
but to have a black pebble on the root at some time.

\begin{figure}
\vspace*{-.5cm}
\hspace*{1.5cm}\includegraphics[scale=0.70]{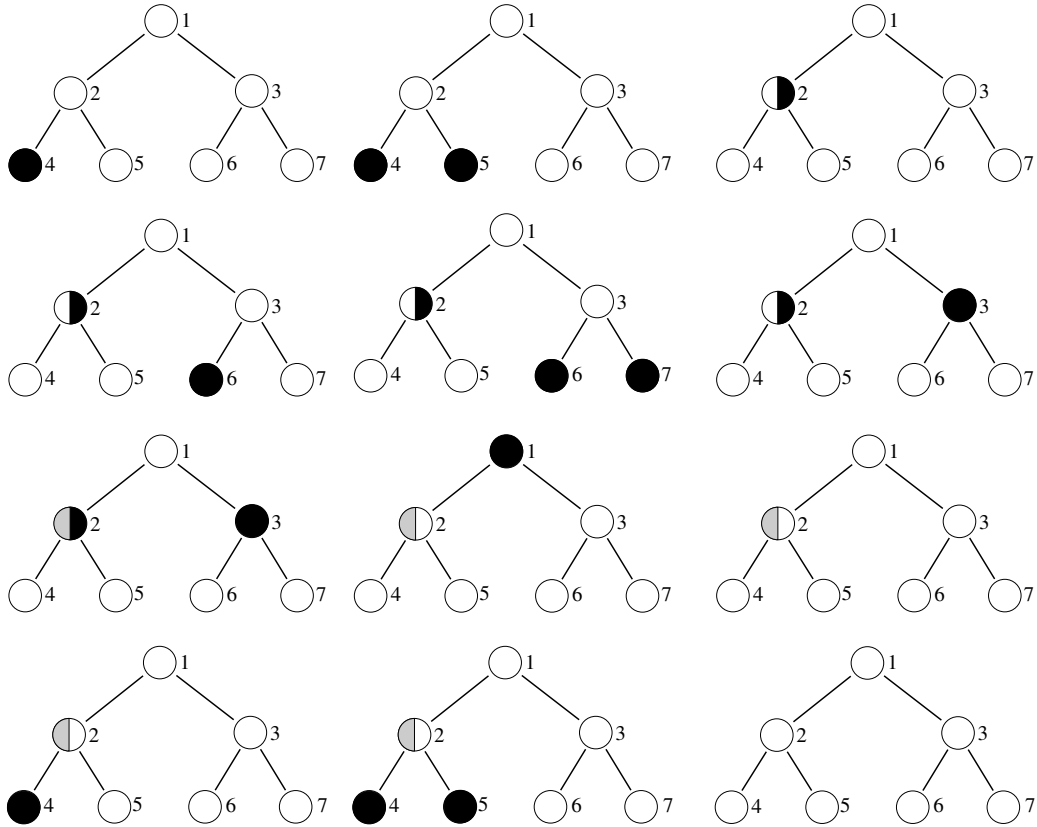}
\vspace*{-1cm}
\caption{An optimal fractional pebbling sequence for the height 3 tree using 2.5 pebbles, all configurations included. The grey half circle means the \emph{white} value of that node is $.5$, whereas unshaded area means absence of pebble value. So for example in the seventh configuration, node 2 has black value .5 and white value .5, node 3 has black value 1, and the remaining nodes all have black and white value 0. }
\label{f:bin_h3_fract_ub}
\end{figure}

The third is {\em fractional pebbling},
which generalises whole black-white pebbling by allowing each node 
$i$ to have a \emph{black value} $b(i)$ and a \emph{white value} $w(i)$ 
such that $b(i) + w(i) \le 1$. The total pebble value (i.e. $b(i) + w(i)$) of each
child of a node $i$ must be 1 before the black value of $i$
is increased or the white value of $i$ is decreased.
Figure \ref{f:bin_h3_fract_ub} shows the sequence of configurations
for an optimal fractional pebbling of the binary tree of
height three using 2.5 pebbles.

Our motivation for choosing these definitions is that we want
pebbling algorithms for trees to closely correspond to $k$-way
branching program algorithms for the tree evaluation problem.
If, as in the survey by Razborov \cite{ra91}, we instead used 
\emph{switching and rectifier networks} instead of nondeterministic
branching programs, where input variable labels are on the edges rather
than the nodes, and a node can have any number of out-edges, and the
size of the program is defined as the number of edges, then 
we would get better upper bounds by using a variant of fractional pebbling
where the following analogue of ``white sliding moves'' are allowed: Suppose you 
want to remove white value from an internal node $i$ by first increasing the white value
of one or both of the children of $i$. With white sliding moves, you can combine
those two moves. A precise definition is given in \cite{manuscript}, where it is
also shown that the height 4 binary tree can be fractionally pebbled using white 
sliding moves with 
$8/3$ pebbles, from which it follows that there are 
switching and rectifier networks with $O(k^{8/3})$ edges that solve $\BT[4][k]$.
In contrast, it is shown in \cite{manuscript} that 3 pebbles are necessary and 
sufficient using our chosen definition of fractional pebbling.

Now we give the formal definition of fractional pebbling, and then define
the other two notions as restrictions on fractional pebbling.

\begin{defn}[Pebbling]
\label{d:pebbling}
A {\em fractional pebble configuration} on a rooted $d$-ary
tree $T$ is an assignment of
a pair of real numbers $(b(i),w(i))$ to each node $i$ of the tree, where
\begin{align}
   &  0\le b(i),w(i) \label{e:consOne} \\ &  b(i)+w(i)\le 1
   \label{e:consTwo}
\end{align}
 Here $b(i)$ and $w(i)$ are the
{\em black pebble value} and the {\em white pebble value}, respectively,
of $i$, and $b(i)+w(i)$ is the {\em pebble value} of $i$.  The number of
pebbles in the configuration is the sum over all nodes $i$ of the pebble
value of $i$.  The legal pebble moves are as follows (always subject to
maintaining the constraints (\ref{e:consOne}), (\ref{e:consTwo})): (i)
For any node $i$, decrease $b(i)$ arbitrarily, (ii) For any node $i$,
increase $w(i)$ arbitrarily, (iii) For every node $i$,
if each child of
$i$ has pebble value 1, then decrease $w(i)$ arbitrarily, increase $b(i)$
arbitrarily, and simultaneously decrease the black pebble values
of the children of $i$ arbitrarily. 
\footnote{It is easy to show that we can require, without increasing the pebbling cost, 
that every type (ii) move to increase $w(i)$ so that $b(i)+w(i)=1$, 
and a type (iii) move to decrease $w(i)$ to 0, but we will not need to use that fact here.}

A {\em fractional pebbling} of $T$ using $p$ pebbles is any
sequence of (fractional) pebbling moves on nodes of $T$ which starts
and ends with every node having pebble value 0, and at some point the
root has black pebble value 1, and no configuration has more than $p$
pebbles.

A {\em whole black-white pebbling} of $T$ is a fractional pebbling of
$T$ such that $b(i)$ and $w(i)$ take values in $\{0,1\}$ for every node
$i$ and every configuration.  A {\em black pebbling} is a
black-white pebbling in which $w(i)$ is always 0.
\end{defn}

Notice that rule (iii) does not quite treat black and white pebbles
dually, since the pebble values of the children must each be 1 before any
decrease of $w(i)$ is allowed.  A true dual move would allow increasing
the white pebble values of the children so they all have pebble value 1
while simultaneously decreasing $w(i)$.  In other words, we allow black
sliding moves, but disallow white sliding moves.  The reason for this
(as mentioned above) is
that non-deterministic branching programs can simulate the former,
but not the latter.

We use $\Bpebbles(T)$, $\BWpebbles(T)$, and $\FRpebbles(T)$ respectively
to denote the minimum number of pebbles required to black pebble $T$,
black-white pebble $T$, and fractional pebble $T$.
Bounds for these values are given in \cite{fsttcs}\footnote{And also for arbitrary degree $d$}.
For example, $\Bpebbles(T^h)= h$,
$\BWpebbles(T^h)= \lceil h/2\rceil +1$, and
$\FRpebbles(T^h) \le h/2+1$ (see \cite{manuscript} for proofs).  
In particular $\FRpebbles(T^3) = 2.5$ (see Figure \ref{f:bin_h3_fract_ub}).

%% file: upperbound_and_height2_and_minDepth.tex
\subsection{Upper Bound for Thrifty BPs}

It is easy to show that the determinstic thrifty BPs we get from pebbling have $O(k^h)$ states, for all $h$. The next theorem shows there is a simple expression for the exact number of states. We do not know how to beat this upper bound for any $k$ and $h$, even by one.
\begin{theorem}
There are $(k+1)^h$ state deterministic thrifty BPs solving $\FT$.
\end{theorem}
\begin{proof}
 
 For $h=1$ you have the start state that queries the single input variable $l_1$, with an edge out to each of the $k$ output states. 

 For $h \ge 2$, we start with $k+1$ copies $B_0,B_1...,B_k$ of the BP that computes $\FT[h-1][k]$. Here is the idea. We will use $B_0$ to compute the value of the left subtree, and for each $a \in [k]$ we use $B_a$ to compute the value of the right subtree while remembering the value of the left subtree. At the level just before the output states, for each $\pair{a,b} \in [k]^2$ there is a state that queries $f_1(a,b)$. 

  Now for the formal definition. We will combine $B_0,B_1,\ldots,B_k$ in such a way that $B_1,...,B_k$ are pairwise disjoint, and for all $a \in [k]$,\ $B_0$ and $B_a$ intersect in exactly one state; namely, for all $a \in [k]$, if $q_{0,a}$ is the output state of $B_0$ labeled $a$, and $q_a$ is the start state of $B_a$, then we remove $q_{0,a}$ and for each of the now-dangling $B_0$-edges $e$, we connect the free end of $e$ to $q_a$.

  Now change the state labels of $B_0$ so that whenever it queries $f_i(b_1,b_2)$ (resp. $l_i$) for some $i \in T^{h-1}$, it instead queries $f_{\sigma_2(i)}(b_1,b_2)$ (resp. $l_{\sigma_2(i)}$) where $\sigma_2$ maps node labels of $T^{h-1}$ to node labels of the subtree of $T^h$ rooted at node 2, in the obvious way. Similarly, for each $a$ in $[k]$, change the state labels of $B_a$ so that whenever it queries $f_i(b_1,b_2)$ (resp. $l_i$) for some $i \in T^{h-1}$, it instead queries the variable $f_{\sigma_3(i)}(b_1,b_2)$ (resp. $l_{\sigma_3(i)}$) where $\sigma_3$ is like $\sigma_2$ except it maps node labels of $T^{h-1}$ to node labels of the subtree of $T^h$ rooted at node 3.

  Next, for each $a,b$ in $[k]$, change the $b$ labeled output state of $B_a$ into a state that queries $f_1(a,b)$. 
  Finally, add in the obvious way (there is only one way) $k$ new output states that receive edges from the $k^2$ former output states of $B_1,...,B_k$. That completes the definition of the BP the computes $\FT[h][k]$. Its size $s(h,k)$ is given by 
  \[ s(h,k) = (k+1) \, s(h-1,k) - k + k = (k+1)^h\]   
  Where the $-k$ is for the states $q_1,\ldots,q_k$ that get counted twice in the expression $(k+1) \, s(h-1,k)$ and the $+k$ is for the new output states.
\end{proof}
%\vspace{15pt}
\subsection{Upper Bound is Exact for Height 2}
We can show the previous upper bound $(k+1)^2$ is the exact state cost $\FT[2][k]$ (note $k+1$ is obviously exact for $\FT[1][k]$). In Section \ref{s:open_problems} we conjecture that $(k+1)^3$ is exact for height 3 as well.

\begin{theorem}
 Every BP solving $\FT[2][k]$ has at least $(k+1)^2$ states.
\end{theorem}
\begin{proof}
There are at least $k^2$ states that query the root, since for all $a,b$ there is at least one state that queries $f_1(a,b)$. There are $k$ output states.

Let $E^*$ be the inputs such that $f_1$ is $+$ mod $k$. Let $Q^*$ be the states $q$ such that $q$ is the last leaf querying state on the computation path of some $I \in E^*$. We can show $Q^*$ has size at least $k$. Let $g$ be the function that maps each input in $E^*$ to its last leaf querying state. Since $|E^*| = k^2$, it suffices to show that $|g^{-1}(q)| \le k$ for every $q$ in $Q^*$. Let $I_{a,b}$ be the unique input in $E^*$ with $\pair{l_2^I,l_3^I} = \pair{a,b}$. Let $q \in Q^*$ be arbitrary. Consider the case that $q$ queries $l_3$ -- the other case is similar. Then it suffices to show that for every $b$, there is at most one $a$ such that $I_{a,b}$ is in $g^{-1}(q)$. Just observe that if two inputs in $E^*$ reach $q$ then they have the same output state, and the label $a'$ of the output state determines the unique $a$ such that $a' = a + b \mod k$.

Now we want to show there is at least one state that queries a leaf and is not in $Q^*$. Since all the inputs in $E^*$ agree on the $f_i$ variables, there is a unique state $q$ that is the first leaf querying state visited by any of them. Because $+$ mod $k$ is a quasigroup, every input in $E^*$ must query $l_2$ and $l_3$ each at least once. So for every $I \in E^*$ there is a leaf querying state on the computation path of $I$ after $q$ that queries a leaf variable. Hence $q \not \in Q^*$. That is $k^2 + k + k + 1 = (k+1)^2$ states total.
\end{proof}

%\vspace{15pt}
\subsection{Minimum-depth BPs are Thrifty}

Let the depth of a deterministic branching program be the maximum number of states visited by any input, with the output state included. The thrifty programs we get from pebbling have depth $2^h$, and it is easy to show that depth $2^h$ is required, regardless of size; just note that Lemma \ref{l:quasiAllThrifty} holds without the depth restriction. In fact, we can show thrifty programs are the \emph{only} fastest determinstic BPs solving $\BT$.

\begin{theorem}
For all $h,k$ every deterministic branching program of depth at most $2^h$ computing $\BT$ (or $\FT$) is thrifty.
\end{theorem}
\begin{proof}
Let $E_0$ be the inputs all of whose internal node functions are quasigroups, and $E_1$ the inputs that query each node exactly once. 
\begin{lemma}\label{l:quasiAllThrifty}
 Every input in $E_0$ queries each of its thrifty variables.
\end{lemma}
\begin{proof}
 Suppose $I \in E_0$ does not query its thrifty $i$ variable. Let $X$ be the thrifty $i$ variable of $I$. For each $a \not = v_i^I$ there is an input $I_a$ identical to $I$ except $X^{I_a} = a$. Define the function $F_i^I : [k] \to [k]$ by 
\[ 
 F_i^I := 
\begin{cases}
\text{identity} & \text{if $i = 1$} \\
f_j^I(F_j^I, v_{2j+1}^I) & \text{if $i = 2j$} \\
f_j^I(v_{2j}^I, F_j^I) & \text{if $i = 2j+1$}
\end{cases}
\]Since $F_i^I$ is a permutation, the root values of the inputs $I_a$ are all different from each other and from $v^I$. If $v_1^I = 1$ then let $J$ be any of the $I_a$, and otherwise let $J$ be the unique $I_a$ such that $v_1^{I_a} = 1$. Then $J \in \BT$ iff $I \not \in \BT$. But their computation paths are the same, a contradiction.
\end{proof}

\begin{lemma}\label{l:quasiThrifty}
Every input in $E_0$ is thrifty (queries only its thrifty variables).
\end{lemma}
\begin{proof}
Because of the depth restriction, if an input queries each of its thrifty variables, then it is thrifty. So this lemma follows from Lemma \ref{l:quasiAllThrifty}.
\end{proof}

\begin{lemma}\label{l:E1Thrifty}
 Every input in $E_1$ is thrifty.
\end{lemma}
\begin{proof}
\def\Xis{X_{i^*}}
\def\fis{f_{i^*}}
 Suppose there is some $I$ in $E_1$ that is not thrifty. For each node $j$, let $X_j$ be the unique $j$ variable that $I$ queries. Since $I$ is not thrifty, there is an internal node $i$ such that $X_i$ is not the thrifty $f_i$ variable of $I$. Let $i^*$ be such a node of minimum height. Since the computation path of $I$ constrains only one value of each internal node function, we can choose an input $J \in E_0$ such that $X_j^I = X_j^J$ for all nodes $j$. $J$ is thrifty by Lemma \ref{l:quasiThrifty}. In particular, $X_{i^*}$ is the thrifty $f_{i^*}$ variable of $J$.
Since $\Xis$ is not the thrifty $\fis$ variable of $I$, it must be that $v_{2i^*}^I \not = v_{2i^*}^J$ or $v_{2i^*+1}^I \not = v_{2i^*+1}^J$. Wlog assume it is the first case. By our choice of $i^*$ and the assumption that $I$ queries every node, we know $I$ queries all its thrifty $T_{2i}$ variables. Since the computation paths of $I$ and $J$ are identical, and $J$ is thrifty, we have that $I$ and $J$ have the same thrifty $T_{2i}$ variables. But then the only way to have $v_{2i^*}^I \not = v_{2i^*}^J$ is if there is a $T_{2i}$ variable $X$ that is thrifty for $I$ (and so also for $J$) such that $X^I \not = X^J$. This contradicts the definition of $J$.
\end{proof}

Let $E_2 := E - E_1$. Fix $I$ in $E_2$. Let $P^I$ be the maximum length initial segment of the computation path of $I$ such that there is some $J$ in $E_0$ for which $P^I$ is also an initial segment of the computation path of $J$. Fix such a $J$. 
Since $I$ is not in $E_1$, there must be some $i$ such that $I$ does not query any $i$ variable. So by Lemma \ref{l:quasiAllThrifty}, we know $P^{I}$ cannot be the entire computation path of $I$ (because then it would be the entire computation path of $J$). So the last state $q_t$ of $P^{I}$ cannot be its output state. Let $q_{t+1}$ be the next state that $I$ visits and $e_t$ the edge $I$ takes from $q_t$ to $q_{t+1}$. Let $X_t$ be the variable queried by $q_t$ and $i_t := \var(X_t)$. 
There must be at least one $J$ in $E_0$ that follows $P^{I}$ (note the definition allows $P^I$ to be a single state). Let $q_{t+1}'$ be the next state visited by $J$. Since $J$ disagrees with $I$ on $X_t$, it must be that $q_t$ is the first state on the computation path of $I$ that queries $X_t$. On the other hand, there must have been a state $q_s$ before $q_t$ on $P^{I}$ that queries an $i_t$ variable $X_s$ distinct from $X_t$; otherwise, there would be a $J'$ in $E_0$ such that $P^{I},e_t,q_{t+1}$ is an initial segment of the computation path of $J'$, contradicting the maximality of $P^I$. So now we know that $J$ queries two distinct $i_t$ variables. But $J$ is in $E_1$ (since $E_0 \subseteq E_1$), so this contradicts Lemma \ref{l:quasiThrifty}.
\end{proof}

%% file: thrifty_advice.tex
\subsection{Lower Bound for Thrifty BPs}

Now we give a tight lower bound for deterministic thrifty BPs. As discussed in section \ref{s:branching_programs}, this improves on an earlier result in \cite{fsttcs}, which gives a lower bound of $\frac{1}{2} k^h$ for all $h$ and all $k > {2^h \choose h-1}$.

\begin{theorem}\label{t:thrifty_advice}
For any $h,k$, every deterministic thrifty branching program solving $\BT$ has at least $k^h$ states.
\end{theorem}

Fix a deterministic thrifty BP $B$ that solves $\BT$. Let $E$ be the inputs to $B$. Let $\Vars$ be the set of $k$-valued input variables (so $|E| = k^{|\Vars|}$). Let $Q$ be the states of $B$. If $i$ is an internal node then the $i$ variables are $f_i(a,b)$ for $a,b \in [k]$, and if $i$ is a leaf node then there is just one $i$ variable $l_i$. We sometimes say ``$f_i$ variable'' just as an in-line reminder that $i$ is an internal node. Let $\var(q)$ be the input variable that $q$ queries. Let $\node$ be the function that maps each variable $X$ to the node $i$ such that $X$ is an $i$ variable, and each state $q$ to $\node(\var(q))$. When it is clear from the context that $q$ is on the computation path of $I$, we just say ``$q$ queries $i$'' instead of ``$q$ queries the thrifty $i$ variable of $I$''.

Fix an input $I$, and let $P$ be its computation path. We will choose $n$ states on $P$ as {\bf critical states} for $I$, one for each node. 
Note that $I$ must visit a state that queries the root (i.e. queries the thrifty root variable of $I$), since otherwise the branching program would make a mistake on an input $J$ that is identical to $I$ except 
$f_1^J(v_2^I,v_3^I) := k - f_1^I(v_2^I,v_3^I)$;
 hence $J \in BT^h_2(k)$ iff $I \not \in BT^h_2(k)$. So, we can choose the root critical state for $I$ to be the last state on $P$ that queries the root. The remainder of the definition relies on the following small lemma:
\begin{lemma}\label{l:basic_thrifty}
 For any $J$ and internal node $i$, if $J$ visits a state $q$ that queries $i$, then for each child $j$ of $i$, there is an earlier state on the computation path of $J$ that queries $j$.
\end{lemma}
\begin{proof}
 Suppose otherwise, and wlog assume the previous statement is false for $j=2i$. For every $a \not = v_{2i}^J$ there is an input $J_a$ that is identical to $J$ except $v_{2i}^{J_a} = a$.
But the computation paths of $J_a$ and $J$ are identical up to $q$, so $J_a$ queries a variable $f_i(a,b)$ such that $b = v_{2i+1}^{J_a}$ and $a \not = v_{2i}^{J_a}$, which contradicts the thrifty assumption.
\end{proof} 
Now we can complete the definition of the critical states of $I$. For $i$ an internal node, if $q$ is the node $i$ critical state for $I$ then the node $2i$ (resp. $2i+1$) critical state for $I$ is the last state on $P$ before $q$ that queries $2i$ (resp. $2i+1$).

Now we assign a pebbling sequence to each state on $P$, such that the set of pebbled nodes in each configuration is a minimal cut of the tree or a subset of some minimal cut (and once it becomes a minimal cut, it remains so), and any two adjacent configurations are either identical, or else the later one follows from the earlier one by a valid pebbling move. This assignment can be described inductively by starting with the last state on $P$ and working backwards. Note that implicitly we will be using the following fact:
\begin{fact}\label{f:basic_crit_state}
For any input $I$, if $j$ is a descendant of $i$ then the node $j$ critical state for $I$ occurs earlier on the computation path of $I$ than the node $i$ critical state for $I$.  
\end{fact}
The pebbling configuration for the output state has just a black pebble on the root. Assume we have defined the pebbling configurations for $q$ and every state following $q$ on $P$, and let $q'$ be the state before $q$ on $P$. If $q'$ is not critical, then we make its pebbling configuration be the same as that of $q$. If $q'$ is critical then it must query a node $i$ that is pebbled in $q$. The pebbling configuration for $q'$ is obtained from the configuration for $q$ by removing the pebble from $i$ and adding pebbles to $2i$ and $2i+1$ (if $i$ is an internal node - otherwise you only remove the pebble from $i$). 

In the above definition of the pebbling configurations, consider the first critical state we define that queries a height 2 node (working backwards -- so the first critical state we define queries the root). We use $r^I$ to denote this state and call it the {\bf supercritical state} of $I$. Since the pebbling configurations up to $r^I$ (again, working backwards) are minimal cuts of the tree, and the children of $\node(r^I)$ are included, it is not hard to see that there must be at least $h$ pebbled nodes. We refer to these nodes as the {\bf bottleneck nodes} of $I$. Define the {\bf bottleneck path} of $r \in R$ to be the path from $\node(r)$ to the root. The bottleneck path of $I \in E$ is the bottleneck path of $r^I$. 
This is the main property of the pebbling sequences that we need: 
\begin{fact}\label{f:basic_peb_seq}
For any input $I$, if non-root node $i$ with parent $j$ is pebbled at a state $q$ on $P^I$, then the node $j$ critical state $q'$ of $I$ occurs later on $P^I$, and there is no state (critical or otherwise) between $q$ and $q'$ on $P^I$ that queries $i$. 
\end{fact}
Let $R$ be the states that are supercritical for at least one input. Let $E_r$ be the inputs with supercritical state $r$. Now we can state the main lemma. 
\begin{lemma}\label{l:thrifty_advice_main_lemma}
 For every $r \in R$, there is an injective function 
from $E_r$ to $[k]^{|\Vars|-h}$. 
\end{lemma}
The lemma gives us that $|E_r| \le k^{|\Vars|-h}$ for every $r \in R$. Since $\{E_r\}_{r\in R}$ is a partition of $E$, there must be at least $|E| / k^{|\Vars|-h} = k^h$ sets in the partition, i.e. there must be at least $k^h$ supercritical states. So the theorem follows from the lemma. 

Fix $r \in R$ and let $D := E_r$. Let $\isc := \node(r)$. Since $r$ is thrifty for every $I$ in $D$, there are values $v_{2\isc}^D$ and $v_{2\isc+1}^D$ such that $v_{2\isc}^I = v_{2\isc}^D$ and $v_{2\isc+1}^I = v_{2\isc+1}^D$ for every $I$ in $D$. We are going to define a procedure $\IntAdv$ that takes as input a $[k]$-string (the advice), tries to interpret it as the code of an input in $D$, and when successful outputs that input. We want to show that for every $I \in D$ we can choose $\adv^I \in [k]^{|\Vars|-h}$ such that $\IntAdv(\adv^I)\DEF = I$.
Of course, choosing $\adv^I$ for each $I$ yields the injective function required to prove the lemma.

During the execution of $\IntAdv$ we maintain a current state $q$, a partial function $v^*$ from nodes to $[k]$, and a set of nodes $\UL$. Once we have added a node to $\UL$, we never remove it, and once we have added $v^*(i) := a$ to the definition of $v^*$, we never change $v^*(i)$. We have reached $q$ by following a \emph{consistent partial computation path} starting from $r$, meaning there is at least one input in $D$ that visits exactly the states and edges that we visited between $r$ and $q$. So initially $q = r$. Intuitively, $v^*(i)\DEF = a$ for some $a$ when we have ``committed'' to interpreting the advice we have read so-far as being the initial segment of \emph{some} complete advice string $\adv^I$ for an input $I$ with $v_i^I = a$. Initially $v^*$ is undefined everywhere. As the procedure goes on, we may often have to use an element of the advice in order to set a value of $v^*$; however, by exploiting the properties of the critical state sequences, for each $I \in D$, when given the complete advice $\adv^I$ for $I$ there will be at least $h$ nodes $\UL^I$ that we ``learn'' without directly using the advice. Such an oppurtunity arises when we visit a state that queries some variable $f_i(b_1,b_2)$ and we have not yet committed to a value for at least one of $v^*(2i)$ or $v^*(2i+1)$ (if both then, we learn two nodes). When this happens, we add that child or children of $i$ to $\UL$ (the {\sf L} stands for ``learned''). So initially $\UL$ is empty. There is a loop in the procedure $\IntAdv$ that iterates until $|\UL| = h$. Note that the children of $\isc$ will be learned immediately. Let $v^*(D)$ be the inputs in $D$ consistent with $v^*$, i.e. $I \in v^*(D)$ iff $I \in D$ and $v_i^I = v^*(i)$ for every $i \in \Dom(v^*)$. 

Following is the complete pseudocode for $\IntAdv$. We also state the most-important of the invariants that are maintained. \\

\noindent
{\bf Procedure} $\IntAdv(\vva \in [k]^*)$:
\begin{algorithmic}[1]
  \STATE $q := r$,\ $\UL := \emptyset$,\ $v^* := \text{undefined everywhere}$.
  \STATE {\bf Loop Invariant:} If $N$ elements of $\vva$ have been used, then $|\Dom(v^*)| = N + |\UL|$.\\
  \WHILE{$|\UL| < h$}
  	  \STATE $i := \node(q)$
        \IF{$i$ is an internal node and $2i \not \in \Dom(v^*)$ or $2i+1 \not \in \Dom(v^*)$}
	        \STATE let $b_1,b_2$ be such that $\var(q) = f_i(b_1,b_2)$.
	        \IF{$2i \not \in \Dom(v^*)$}
	        	\STATE $v^*(2i) := b_1$ and $\UL := \UL + 2i$.
		  \ENDIF
		  \IF{$2i+1 \not \in \Dom(v^*)$ and $|\UL| < h$}
		  	\STATE $v^*(2i+1) := b_2$ and $\UL := \UL + (2i+1)$.
	        \ENDIF
        \ENDIF
        \IF{$i \not \in \Dom(v^*)$}
        	\STATE let $a$ be the next unused element of $\vva$.
	       \STATE $v^*(i) := a$.
        \ENDIF
        \STATE $q := $ the state reached by taking the edge out of $q$ labeled $v^*(i)$.
  \ENDWHILE
  \STATE let $\vvb$ be the next $|\Vars| - |\Dom(v^*)|$ unused elements of $\vva$. \label{line:lastadviceuse}
  \STATE let $I_1,\ldots,I_{|v^*(D)|}$ be the inputs in $v^*(D)$ sorted according to some globally fixed order on $E$. \label{line:secondtolastline}
  \STATE if $\vvb$ is the $t$-largest string in the lexiocgraphical ordering of $[k]^{|\Vars| - |\Dom(v^*)|}$, and $t \le |v^*(D)|$, then return $I_t$.\footnote{See after this code for argument that $|v^*(D)| \le k^{|\Vars| - |\Dom(v^*)|}$.}  \label{line:lastline}
\end{algorithmic}

\vspace{10pt}

If the loop finishes, then there are at most $|E|/|\Dom(v^*)| = k^{|\Vars|-|\Dom(v^*)|}$ inputs in $v^*(D)$. 
So for each of the 
inputs $I$ enumerated on line \ref{line:secondtolastline}, there is a way of setting $\vva$ so that $I$ will be chosen on line \ref{line:lastline}. 

Recall we are trying to show that for every $I$ in $D$ there is a string $\adv^I  \in [k]^{|\Vars|-h}$ such that $\IntAdv(\vec{a})\DEF = I$. This is easy to see under the assumption that there is such a string that makes the loop finish while maintaining the loop invariant; since the loop invariant ensures we have used $|\Dom(v^*)| - h$ elements of advice when we reach line \ref{line:lastadviceuse}, and since line \ref{line:lastadviceuse} is the last time when the advice is used, in all we use at most $|\Vars| - h$ elements of advice.
 To remove that assumption, first observe that for each $I$, we can set the advice to some $\adv^I$ so that $I \in g(D)$ is maintained when $\IntAdv$ is run on $\vec{a}^I$. Moreover, for that $\adv^I$, we will never use an element of advice to set the value of a bottleneck node of $I$, and $I$ has at least $h$ bottleneck nodes. Note, however, that this does not necessarily imply that $\UL^I$ (the $h$ nodes $\UL$ we obtain when running $\IntAdv$ on $\adv^I$) is a subset of the bottleneck nodes of $I$. Finally, note that we are of course implicitly using the fact that no advice elements are ``wasted''; each is used to set a different node value.

\begin{cor}
 For any $h,k$, every deterministic thrifty branching program solving $\BT[h][k]$ has at least $\sum_{2 \le l \le h} k^l$ states.
\end{cor}
\begin{proof}
 The previous theorem only counts states that query height 2 nodes. The same proof is easily adapted to show there are at least $k^{h-l+2}$ states that query height $l$ nodes, for $l = 2,\ldots,h$. Those $h-1$ state sets are disjoint, so we can sum the bounds.
\end{proof}

%% file: fractional_lower_bound.tex
The proof of Theorem \ref{t:fractLB} proceeds by reducing the problem of proving lower 
bounds on the fractional pebbling cost for balanced binary trees, to the problem of
proving lower bounds on the black-white pebbling costs for a family of DAGs. In doing
so, we are essentially discretizing the fractional pebbling problem; 
the main construction has a parameter $c$ that determines how many nodes in the dag 
are used to ``simulate'' each node in the tree. We will use the next lemma (due to S. Cook) to 
conclude that we can always make $c$ large enough that we don't ``lose anything''.
\begin{lemma}\label{l:rational} For every finite DAG there is an optimal
fractional B/W pebbling in which all pebble values are rational numbers.
(This result is robust independent of various definitions of pebbling;
for example with or without sliding moves, and whether or not we require
the root to end up pebbled.) 
\end{lemma}
\begin{proof}  Consider an optimal B/W fractional pebbling algorithm.
Let the variables $b_{v,t}$ and $w_{v,t}$ stand for the black and white
pebble values of node $v$ at step $t$ of the algorithm.

{\bf Claim:}  We can define a set of linear inequalities with 0 -
1 coefficients  which suffice to ensure that the pebbling is legal.

For example,  all variables are non-negative, $b_{v,t} + w_{b,t} \le 1$,
initially all variables are 0, and finally the nodes have the values
that we want, node values remain the same on steps in which nothing is
added or subtracted, and if the black value of a node is increased at
a step then all its children must be 1 in the previous step, etc.

Now let $p$ be a new variable representing the maximum pebble value of
the algorithm.  We add an inequality for each step $t$ that says the
sum of all pebble values at step $t$ is at most $p$.

Any solution to the linear programming problem:

    Minimize $p$ subject to all of the above inequalities

gives an optimal pebbling algorithm for the graph.  But Every LP program
with rational coefficients has a rational optimal solution (if it has
any optimal solution). 
\end{proof}

Now we are ready to prove the lower bound. We know this bound is not tight for heights at 
most 4. This is easy to see for height 2 (the bound should be $d$, but the theorem gives $d/2 - 1$), 
and proofs of the tight bounds for heights 3 and 4 are given in \cite{manuscript}.  
\begin{theorem}\label{t:fractLB}
 The fractional pebbling cost for the degree $d$, height $h$ tree is at least $(d-1)h/2 - d/2$.
\end{theorem}

\begin{proof} 

The high-level strategy for the proof is as follows. Given $d$ and $h$, we 
transform the tree $T_{d}^{h}$ into a DAG $G_{d,h}$ such that a lower
bound on $\BWpebbles(G_{d,h})$ gives a lower
bound for $\FRpebbles(T_{d}^{h})$. 
To analyze $\BWpebbles(G_{d,h})$, we use a 
result of
Klawe \cite{klawe}, who shows that for a DAG $G$ that satisfies a certain 
``niceness'' property, $\BWpebbles(G)$ can be given in terms of $\Bpebbles(G)$ 
(and the relationship is tight to within a constant less than one). 
The black pebbling cost is typically
easier to analyze. In our case, $G_{d,h}$ does not satisfy the niceness
property as-is, but just by removing some edges from $G_{d,h}$, we get
a new DAG $G'_{d,h}$ which is nice. We then show how to exactly compute
$\Bpebbles(G'_{d,h})$ which yields a lower bound on
$\BWpebbles(G_{d,h})$, and hence on $\FRpebbles(T_{d}^{h})$.

We first motivate the construction $G_{d,h}$ and show that the whole black-white
pebbling number of $G_{d,h}$ is related to the fractional pebbling number
of $T_{d}^{h}$. 
 
We first use Lemma \ref{l:rational} to ``discretize'' the
fractional pebble game. The following are the rules for the
discretized game, where $c$ is a parameter:
\begin{packed_item}
 \item For any node $v$, decrease $b(v)$ or increase $w(v)$ by $1/c$.
 \item For any node $v$, including leaf nodes, if all the children of $v$
 have value 1, then increase $b(v)$ or decrease $w(v)$ by $1/c$.
\end{packed_item}

By Lemma \ref{l:rational}, we can assume all pebble
values are rational, and if we choose $c$ large enough it is not a restriction
that pebble values can only be changed by $1/c$. Since sliding moves
are not allowed, the pebbling cost for this game is at most one more
than the cost of fractional pebbling with black sliding moves.

Now we show how to construct $G_{d,h}$ (for an example, see figure \ref{f:reductionG}).
 We will split up each node of
$T_{d}^{h}$ into $c$ nodes, so that the discretized game corresponds to
the whole black-white pebble game on the new graph. Specifically, the cost
of the whole black-white pebble game on the new graph will be exactly $c$
times the cost of the discretized game on $T_{d}^{h}$.

In place of each node $v$ of $T_{d}^{h}$, $G_{d,h}$ has $c$ nodes $v[1], \ldots, v[c]$;
having $c'$ of the $v[i]$ pebbled simulates $v$ having value $c'/c$. In
place of each edge $(u,v)$ of $T_{d}^{h}$ is a copy of the complete bipartite 
graph $(U,V)$, where $U$ contains nodes $u[1] \ldots u[c]$ and $V$ contains nodes $v[1]
 \ldots v[c]$. If $u$ was a parent of $v$ in the tree, then all the edges go
from $V$ to $U$ in the corresponding complete bipartite graph. Finally, a new 
``root'' is added at height $h+1$ with edges from each
of the $c$ nodes at height $h$\footnote{The reason for this is quite technical: Klawe's definition 
of pebbling is slightly different from ours in that it requires that the root remain pebbled. Adding 
a new root forces there to be a time when all $c$ of the height $h$ nodes, which represent the
root of $T_d^h$, are pebbled. Adding one more pebble to $G_{d,h}$ changes the relationship between 
the cost of pebbling $T_d^h$ and the cost of pebbling $G_{d,h}$ by a negligible amount.}. 
So every node at height $h-1$ and lower
has $c$ parents, and every internal node except for the root has $dc$
children. 

\begin{figure}
\vspace*{.3cm}
\hspace*{1.5cm}\includegraphics[scale=0.70]{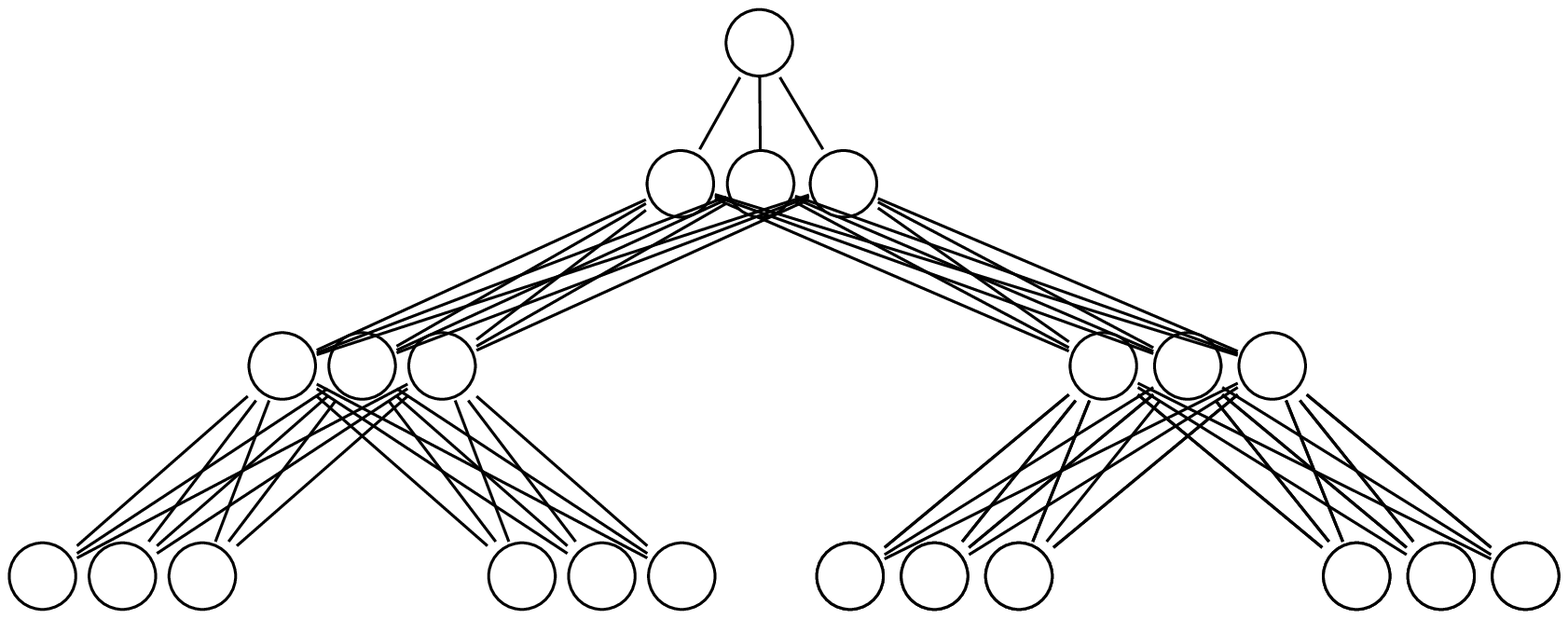}
\hspace*{1cm} \vspace*{.2cm}
\caption{Illustration to accompany the definition of $G_{d,h}$. This is $G_{2,3}$ with parameter $c=3$}
\label{f:reductionG}
\end{figure}

To lower bound $\BWpebbles(G_{d,h})$, we will use
Klawe's result \cite{klawe}. Klawe showed that for ``nice'' graphs $G$, the black-white
pebbling cost of $G$ (with black and white sliding moves) is at least $\lfloor \Bpebbles/2 
\rfloor + 1$. Of course, the black-white
pebbling cost without sliding moves is at least the cost with them. We define 
what it means
for a graph to be nice in Klawe's sense.

\begin{defn}
\label{d:nice}
A DAG $G$ is nice if the following conditions hold:
\begin{enumerate}
\item If $u_1$, $u_2$ and $u$ are nodes of $G$ such that $u_1$ and $u_2$
are children of $u$ (i.e., there are edges from $u_1$ and $u_2$ to $u$),
then the cost of black pebbling $u_1$ is equal to the cost of black
pebbling $u_2$
\item If  $u_1$ and $u_2$ are children of $u$, then there is no path from  $u_1$ to $u_2$ or from $u_2$ to $u_1$.  
\item If $u, u_1, \ldots, u_m$ are nodes none of which has
 a path to another, then there are node-disjoint paths $P_1, \ldots,
 P_m$ such that $P_i$ is a path from a leaf (a node with in-degree 0) to $u_i$ 
and there is no path between $u$ and any node in $P_i$.
\end{enumerate}
\end{defn}

$G_{d,h}$ is not nice in Klawe's sense. We will delete some edges from
$G_{d,h}$ to produce a nice graph $G'_{d,h}$ and we will analyze $\Bpebbles(G'_{d,h})$. 
Note that a lower bound on $\BWpebbles(G'_{d,h})$ is also a lower bound on $\BWpebbles(G_{d,h})$.

The following definition will help in explaining the construction of
$G'_{d,h}$ as well as for specifying and proving properties of certain paths. 

\begin{defn} For $u \in G_{d,h}$, let $T_{d}^{h}(u)$ be the node in $T_{d}^{h}$ such that
$T_{d}^{h}(u)[i] = u$ for some $i \leq c$.  For $v,v' \in T_{d}^{h}$, we say $v < v'$
if $v$ is visited before $v'$ in an inorder traversal of $T_{d}^{h}$. For $u,u'
\in G_{d,h}$, we say $u < u'$ if $T_{d}^{h}(u) < T_{d}^{h}(u')$ or if for some $v \in T_{d}^{h}$, $u =
v[i]$, $u' = v[j]$, and $i < j$. 
\end{defn}

$G_{d,h}'$ is obtained from $G_{d,h}$ by removing $c-1$ edges from each internal node
except the root, as follows (for an example, see figure \ref{f:reductionGprime}). 
For each internal node $v$ of $T$, consider
the corresponding nodes $v[1], v[2], \ldots, v[c]$ of $G_{d,h}$. Remove the
edges from $v[i]$ to its $i-1$ smallest and $c-i$ largest children. So
in the end each internal node except the root has $c(d-1)+1$ children.

\begin{figure}
\vspace*{.3cm}
\hspace*{1.5cm}\includegraphics[scale=0.70]{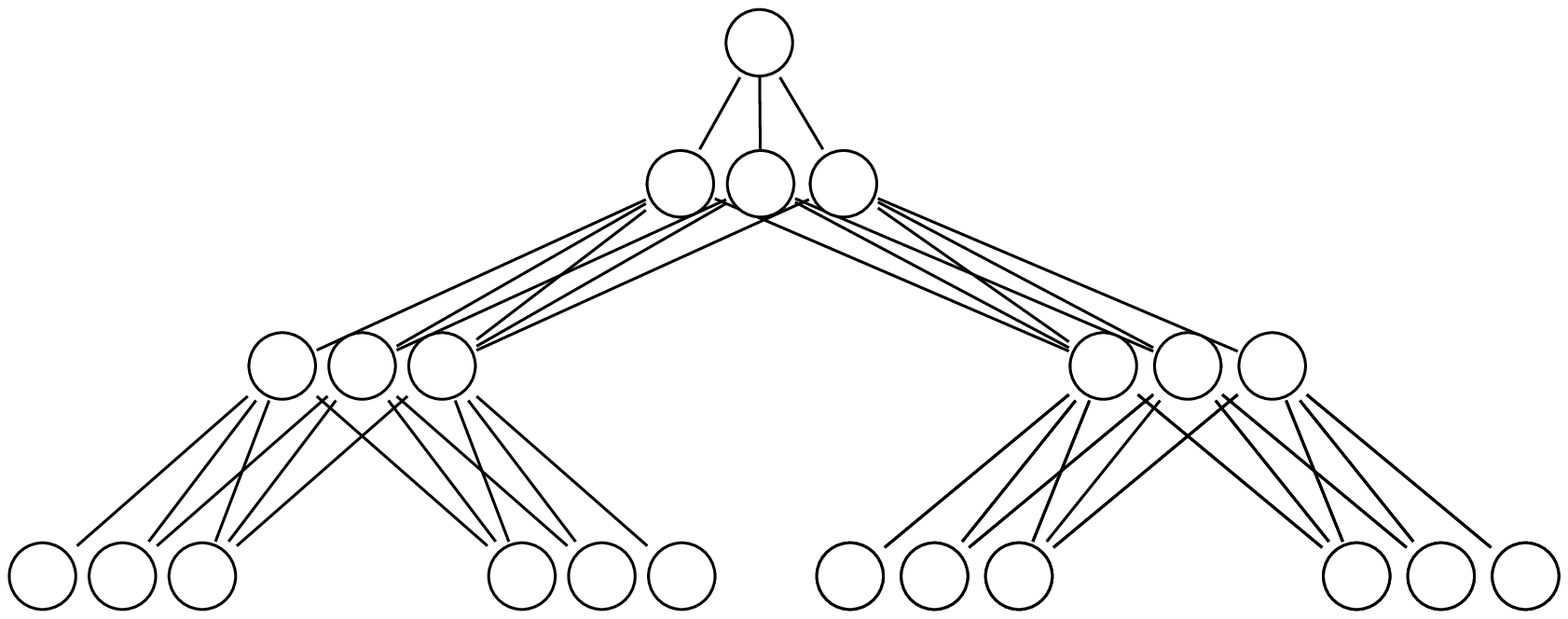}
\hspace*{1cm} \vspace*{.2cm}
\caption{Illustration to accompany the definition of $G'_{d,h}$. This is $G'_{2,3}$ with parameter $c=3$}
\label{f:reductionGprime}
\end{figure}

We first analyze \Bpebbles($G'_{d,h})$ and then show that
it is nice.
We show that $\Bpebbles(G_{d,h}') = c[(d-1)(h-1) +
1]$. Note that an upper bound of $c[(d-1)(h-1) +
1]$ is attained using a simple recursive algorithm similar to that used
for the binary tree.

For the lower bound, consider the earliest time $t$ when all paths from
 a leaf to the root are blocked. Figure \ref{f:fract_bottleneck} 
 is an example of the type of
 pebbling configuration that we are about to analyze. 
 The last pebble placed must have been
placed at a leaf, since otherwise $t-1$ would be an earlier time when
all paths from a leaf to the root are blocked. Let $P$ be the
newly-blocked path from a leaf to the root.  Consider the set $S = \{
u \in G_{d,h}' \ \vert\ u \not \in P \text{ and $u$ is a child of a node in }
P \}$ of size $c (d-1)(h-1) + (c-1) = c[(d-1)(h-1) + 1] - 1$ (the $c-1$
is contributed by nodes at height $h$).  We will give a set of
mutually node-disjoint paths $\{P_u\}_{u \in S}$ such that $P_u$ is a
path from a leaf to $u$ and $P_u$ does not intersect $P$. At time $t-1$,
there must be at least one pebble on each $P_u$, since otherwise there would
still be an open path from a leaf to the root at time $t$. Also counting
the leaf node that is pebbled at $t$ gives c[(d-1)(h-1) + 1] pebbles.

\begin{defn}
 The left-most (right-most) path to $u$ is the unique path ending at
 $u$ determined by choosing the smallest (largest) child at every level.
\end{defn}
\begin{defn}
 $P(l)$ is the node of path $P$ at height $l$, if it exists.
\end{defn}

For each $u \in S$ at height $l$, if $u$ is less than (greater than)
$P(l)$ then make $P_u$ the left-most (right-most) path to $u$. Now
we need to show that the paths $\{P_u\}_{u \in S} \cup \{P\}$ are
disjoint. The following fact is clear from the definition of $G_{d,h}'$.
\begin{lemma}\label{l:thefact} For any $u,u' \in G_{d,h}'$, if $u < u'$ then the
smallest child of $u$ is not a child of $u'$, and the largest child of
$u'$ is not a child of $u$. 
\end{lemma}

First we show that $P_u$ and $P$ are disjoint. The following lemma will
help now and in the proof that $G'_{d,h}$ is nice.

\begin{lemma}\label{l:paths}
 For $u,v \in G_{d,h}'$ with $u < v$, if there is no path from $u$ to $v$
 or from $v$ to $u$ then the left-most path to $u$ does not intersect
 any path to $v$ from a leaf, and the right-most path to $v$ does not
 intersect any path to $u$ from a leaf.
\end{lemma}
\begin{proof} Suppose otherwise and let $P_u'$ be the left-most
path to $u$, and $P_v'$ a path to $v$ that intersects $P_u'$. Since
there is no path between $u$ and $v$, there is a height $l$, one
greater than the height where the two paths first intersect, such that
$P_u'(l), P_v'(l)$ are defined and $P_u'(l) < P_v'(l)$. But then from
Lemma \ref{l:thefact} $P_u'(l-1) \not = P_v'(l-1)$, a contradiction. The
proof for the second part of the lemma is similar. 
\end{proof}

That $P_u$ and $P$ are disjoint follows from using Lemma \ref{l:paths}
on $u$ and the sibling of $u$ in $P$. 

Next we show that for distinct
$u,u' \in S$, $P_u$ does not contain $u'$. Suppose it does. Assume $P_u$
is the left-most path to $u$ (the other case is similar). Since $u
\not = u'$, there must be a height $l$ such that $P_u(l)$ is defined
and $P_u(l-1) = u'$. From the definition of $S$, we know $P(l)$ is also
a parent of $u'$. From the construction of $P_u$, since we assumed $P_u$
is the left-most path to $u$, it must be that $P_u(l) < P(l)$. But then
Lemma \ref{l:thefact} tells us that $u'$ cannot be a child of $P(l)$, a
contradiction. 

The proof that $P_u$ and $P_{u'}$ do not intersect is by contradiction.
Assuming that there are $u,u' \in S$ such that
$P_u$ and $P_{u'}$ intersect, there is a height $l$, one greater
than the height where they first intersect, such that $P_u(l) \not =
P_{u'}(l)$. Note that $P_u$ and $P_{u'}$ are both left-most paths or both
right-most paths, since otherwise in order for them to intersect they
would need to cross $P$. But then from Lemma \ref{l:thefact} $P_u(l-1)
\not = P_{u'}(l-1)$, a contradiction.

See Figure \ref{f:fract_bottleneck} for an example of a bottleneck of the specified structure for $G_{d,h}'$
corresponding to the height 3 binary tree, with $c=3$:

\begin{figure}
\vspace*{.3cm}
\hspace*{1.5cm}\includegraphics[scale=0.70]{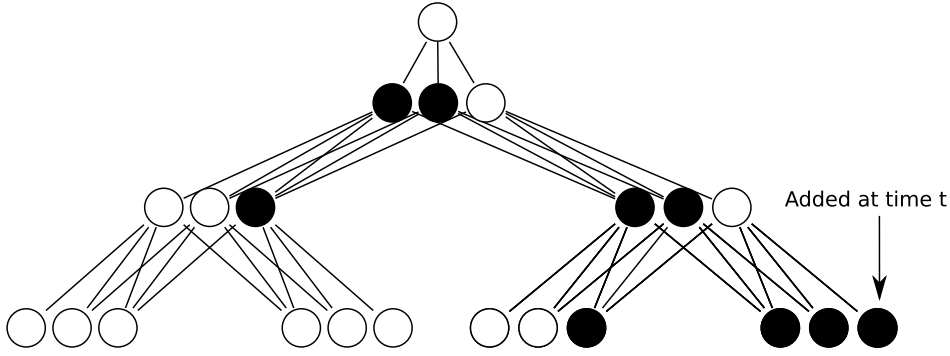}
\hspace*{1cm} \vspace*{.4cm}
\caption{A possible black pebbling bottleneck of $G_{2,3}'$, with $c=3$}
\label{f:fract_bottleneck}
\end{figure}

The last step is to prove that $G_{d,h}'$ is nice. There are three
properties specified in Definition \ref{d:nice}.
Property 2 is obviously satisfied. For property 1, the argument used to
give the black pebbling lower bound of $c[(d-1)(h-1) + 1]$ can be used
to give a black pebbling lower bound of $c(d-1)(l-1) + 1$ for any node
at height $l \leq h$ (the 1 is for the last node pebbled, and recall
the root is at height $h+1$), and that bound is tight. For property 3,
choose $P_i$ to be the left-most (right-most) path from $u_i$ if $u_i$
is less than (greater than) $u$. Then use Lemma \ref{l:paths} on each
pair of nodes in $\{u, u_1,\ldots, u_m\}$. 

Since $\Bpebbles(G'_{d,h}) = c[(d-1)(h-1)+ 1]$, we have
\[\BWpebbles(G_{d,h}) \geq \BWpebbles(G'_{d,h}) \geq c[(d-1)(h-1) + 1]/2\] and 
thus that the pebbling cost for the discretized game on $T_{d}^{h}$ is
at least $(d-1)(h-1)/2 + .5$, which implies $\FRpebbles(T_{d}^{h}) \geq
(d-1)(h-1)/2 - .5$.
\end{proof}

%% file: less_thrifty.tex
\subsubsection{Thrifty BPs with Wrong-Wrong Queries}

A variable $f_i(a,b)$ is {\bf wrong-wrong} for input $I$ iff $a \not = v_{2i}^I$ and $b \not = v_{2i+1}^I$. The next theorem shows that querying wrong-wrong variables does not help. 

\begin{theorem}\label{t:less_thrifty}
For any $h,k \ge 2$, if $B$ is a deterministic BP that solves $\BT$ such that each input only queries variables that are thrifty or wrong-wrong for it, then $B$ has at least $k^h$ states.
\end{theorem}
\begin{proof}

We use the definitions and conventions introduced in the first paragraph of the proof of Theorem \ref{t:thrifty_advice}. The proof of the following lemma is similar to that of Lemma \ref{l:basic_thrifty} (page \pageref{l:basic_thrifty})\footnote{Also this lemma is proved in a more-general context on page \pageref{l:basic_param_less_thrifty}}:
\begin{lemma}\label{l:basic_less_thrifty}
  For any $J$ and internal node $i$, there is at least one state $q$ on the computation path of $J$ that queries the thrifty $i$ variable of $J$, and for every such $q$, for each child $j$ of $i$, there is a state on the computation path of $J$ before $q$ that queries the thrifty $j$ variable of $J$.
\end{lemma}

Recall that for the thrifty lower bound, to each input we assigned one ``critical state'' for each node, and a pebbling configuration to each critical state, such that the $n$ pebbling configurations made a valid pebbling sequence. This was so even if the thrifty branching program was constructed based on a pebbling sequence of length greater than $n$. Now we will not be selecting critical states, and we will assign pebbling sequences with length possibly greater than $n$. It may be helpful to note that this way of assigning pebbling sequences will have the following property:
\def\footfamily{We are talking about a particular family of thrifty BPs $\{B_{S,k}\}$, without taking the time to give a precise definition. $B_{S,k}$ has $|S|$ non-output layers (where $|S|$ is the number of moves in $S$), and if a pebble is placed on $i$ in the $l$-th move of $S$ when there are $p$ pebbles on the tree, then there are $k^p$ states in layer $l$ of $B_{S,k}$, all of which query a node $i$ variable.} 
\begin{remark}
Let $S$ be a complete pebbling sequence for $T^h$ such that the root is pebbled only once, and a pebble is removed from a non-root node $i$ only during a move that places a pebble on the parent of $i$. For any $k$, if $B_{S,k}$ is the thrifty deterministic BP for solving $\FT$ that implements $S$ in the natural way\footnote{\footfamily}, then for every input $I$ to $B_{S,k}$, we will assign pebbling sequence $S$ to $I$.
\end{remark}
In the end, this will result in a cleaner proof; in particular, we will be able to say that when we interpret the advice for $I$, every node that gets ``learned'' is a bottleneck node of $I$ (see Fact \ref{f:learnBnNodes}).

We define the pebbling sequence for $I \in E$ by following the computation path of $I$ from beginning to end, associating the $t$-th thrifty state $q_t$ visited by $I$ with the $t$-th pebbling configuration $C_t$, such that $C_{t+1}$ is either identical to $C_t$ or follows from $C_t$ by applying a valid pebbling move. There is also a last pebbling configuration that is not associated with any state. Let $q_1,\ldots,q_{t^*}$ be the thrifty states on the computation path of $I$, up to the first state $q_{t^*}$ that queries the thrifty root variable of $I$. Note that $q_1$ must query a leaf by Lemma \ref{l:basic_less_thrifty}. We associate $q_1$ with the empty configuration $C_1$.

Assume we have defined the configurations $C_1,\ldots,C_t$ associated with the first $t \le t^*$ thrifty states, and assume $C_1,\ldots,C_t$ is a valid sequence of configurations (where adjacent identical configurations are allowed), but neither it nor any prefix of it is a complete pebbling sequence. We also maintain that for all $t' \le t$, if $\node(q_{t'})$ is internal, then its children are pebbled in $C_{t'}$ and it is not. Let $i := \node(q_{t})$. By the I.H. $i$ is not pebbled in $C_t$. We define $C_{t+1}$ by saying how to obtain it by modifying $C_{t}$:
\begin{ppe} 
  \item If $i$ is the root, then clearly $t=t^*$, and by the I.H. nodes 2 and 3 are pebbled. Put a pebble on the root and remove the pebbles from nodes 2 and 3. This completes the definition of the pebbling sequence for $I$. 
  \item If $i$ is a non-root internal node, then by the I.H. both children of $i$ are pebbled. 
   For each child $j$ of $i$: if there is a state $q'$ after $q_t$ that queries the thrifty $i$ variable of $I$, and no state between $q_{t}$ and $q'$ that queries the thrifty $j$ variable of $I$, then leave the pebble on $j$, and otherwise remove it.
  \item If $i$ is not the root, then place a pebble on $i$ iff there is a state $q'$ after $q_t$ that queries the thrifty $\Par(i)$ variable of $I$ and there is no state between $q_{t}$ and $q'$ that queries the thrifty $i$ variable of $I$.
\end{ppe}

Now we use the classic argument that $h$ pebbles are required to black pebble $T^h$. The children of the root are pebbled in $C_{t^*}$, so $C_{t^*}$ trivially has the property that there is at least one node blocking every path from the root to a leaf. So consider the first $t \le t^*$ such that $C_t$ has that property. Then $i := \node(q_{t-1})$ must be a leaf; otherwise there would be an earlier configuration with the aforementioned property. Consider the first $t' \ge t$ such that $q_{t'}$ queries the thrifty $\Par(i)$ variable of $I$; such a state must exist by the definition of the pebbling sequence for $I$. Then we make $r^I := q_{t'}$ be the {\bf supercritical state} of $I$. We refer to the nodes pebbled in $C_{t'}$ as the {\bf bottleneck nodes} of $I$. Let $R$ be the states that are supercritical for at least one input, and for each $r \in R$ let $E_r$ be the inputs with supercritical state $r$. For $r \in R$ we write $\isc^r$ for $\node(r)$, and for $I \in E_r$ we refer to $\isc^r$ as the {\underline {\sf s}}uper{\underline {\sf c}}ritical node for $I$. 

The definition of the {\bf bottleneck path} $\BnPath_r$ for $r \in R$ has not changed: it is the path from $\isc^r$ to the root.
We mentioned earlier that every node we ``learn'' for an input $I$ is a bottleneck node of $I$. This is due to the next fact. For any $I$ and $q$ on the computation path of $I$, let $\rPath[I][q]$ be the part of the computation path of $I$ starting with $q$. 
\begin{fact}\label{f:learnBnNodes}
 $i$ is a bottleneck node of $I \in E_r$ {\bf iff} it is not in $\BnPath_r$ and there is a state $q \in \rPath[I]$ that queries the thrifty $\Par(i)$ variable of $i$ and no state before $q$ in $\rPath[I]$ that queries the thrifty $i$ variable of $I$.
\end{fact} 

It will be convenient to have named the following four sets of nodes: 
\vspace{3pt}
\begin{defn}[$\SiblBnPath_r, \RightPath_i, \Learnable_r, \Learnable_r^*$] \label{d:NodeSets}
\begin{ppi}
 \item[] {}
 \item $\SiblBnPath_r$ is the set of nodes that are the sibling of a node in $\BnPath_r$. 
 \item For $i \in \SiblBnPath_r$,\ $\RightPath_i$ is the path from $i$ to the right-most leaf under $i$ (when the tree is drawn in the canonical way).
 \item $\Learnable_r$ is the set of nodes $\{2\isc^r, 2\isc^r+1 \} + \bigcup_{i \in \SiblBnPath_r} \RightPath_i$, i.e. the nodes not on the bottleneck path that are the descendent of a node on the bottleneck path.
 \item $\Learnable_r^* := \Learnable_r - \{2\isc^r,2\isc^r+1\}$.
\end{ppi}
\end{defn} \noindent It is not hard to see that every $I \in E_r$ has at least one bottleneck node in $\RightPath_j$ for each of the $h-2$ nodes $j \in \SiblBnPath_r$. % (this observation is used in the black pebbling lower bound argument).
Additionally both children of $\isc^r$ are always bottleneck nodes of $I$, so $I$ has at least $h$ bottleneck nodes. 

Let $G$ be the set of partial functions from $\Vars$ to $[k]$. At least when $k=2$ these are commonly called \emph{restrictions} (of $\BT$), so we will refer to them as restrictions. For $g \in G$ and $D \subseteq E$ we write $g(D)$ for the inputs in $D$ consistent with $g$ -- i.e. $g(D) := \{ I \in D \ | \ \forall X \in \Dom(g).\ X^I = g(X) \}$.
It will be convenient to further partition the sets $E_r$ by fixing some of the variables initially. This finer partitioning 
 appears in the statement of the main lemma:

 \begin{lemma}[Main Lemma]\label{l:less_thrifty_main_one}
 For some integer $M$, for every supercritical state $r \in R$, there is a set of restrictions $\Ginit{r}$ of size at most $k^{|\Vars| - M}$ such that $\{\ginit(E_r) \}_{\ginit \in \Ginit{r}}$ is a partition of $E_r$ and for every $\ginit$ in $\Ginit{r}$, there is an injective function from $\ginit(E_r)$ to $[k]^{M-h}$.
 \end{lemma}

Let us see why the theorem follows from the lemma. Since $\{\ginit(E_r)\}_{\ginit \in \Ginit{r}}$ is a partition of $E_r$, and $\Ginit{r}$ has size at most $k^{|\Vars| - M}$, there must be some $\ginit^* \in \Ginit{r}$ such that $\ginit^*(E_r)$ has size at least $|E_r| / k^{|\Vars| - M}$. On the other hand, from the lemma we get that \emph{every} set $\ginit(E_r)$ in the partition has size at most $k^{M-h}$.
 Hence  \[|E_r| / k^{|\Vars| - M} \le |\ginit^*(E_r)| \le k^{M-h} \]
 Rearranging gives $|E_r| \le k^{|\Vars|} / k^h = |E| / k^h$, and this holds for all $r \in R$. Since $\{E_r\}_{r \in R}$ is a partition of $E$, we get that $R$ must have size at least $k^h$.
 
 \subsubsection*{Proof of Main Lemma}
 
 \newcommand{\preG}{\hat{G}}
  
 \indent We use $T$ to refer to the height $h$ balanced binary tree, or to the set of its nodes. We use $T_i$ to refer to the subtree of $T$ rooted at node $i$, or to its nodes. For $U$ a set of nodes, $\Vars(U)$ is the set of input variables corresponding to the nodes in $U$ -- i.e $\Vars(U) := \{ X \in \Vars \ | \ X = l_i \text{ or } X = f_i(a,b) \text{ for some } i \in U \text{ and } a,b \in [k] \}$. For $D \subseteq E$ there is a partial function $i \mapsto v_i^{D}$ from $T$ to $[k]$  such that $v_i^D\DEF = a$ iff $v_i^I = a$ for every $I$ in $D$. Similarly there is a partial function $X \mapsto v_i^{D}$ from $\Vars$ to $[k]$ such that $X^{D}\DEF = a$ iff $X^I = a$ for every $I$ in $D$. 

The constant $M$ mentioned in the theorem is $k (h-1)(h-2)/2 + k^2(h-1) + h$, but we are just writing that expression here for clarity; we will not be reasoning about it. For each $r \in R$, we are going to define a set $\Ginit{r}$ of at most $k^{|\Vars| - M}$ restrictions where each $\ginit \in \Ginit{r}$ is defined on some set of $|\Vars| - M$ variables. Before giving the precise definition of the partition, let us see where the expression for $M$ comes from. For $(h-1)(h-2)/2 = (h-2) + (h-3) + ... + 1$ internal nodes $i$ we will fix all but $k$ of the $k^2$ variables that define the corresponding function $f_i$. For each of the $h-1$ nodes on the bottleneck path $\BnPath_r$, we will not fix any of the $k^2$ variables that define the corresponding function. Lastly, there will be $h$ unfixed leaf variables. 

Let $\Ufixed^r$ be all the nodes except $\Learnable_r + \BnPath_r$. 
In the following drawing, which depicts the construction for the height 5 tree when $\isc^r = 15$ is the right-most height 2 node, the pruned nodes (the nodes in the subtrees that would be at the ends of the dashed lines) are $\Ufixed^r$ and the unmarked nodes plus the $\triangle$-marked nodes are $\Learnable_r$. 
The $\square$-marked nodes are $\BnPath_r$ and will have no fixed variables. The $\triangle$-marked nodes are $\SiblBnPath_r$ and will have $k^2-k$ fixed variables. 

\vspace{.3cm}
\hspace{4cm}\includegraphics[scale=0.4]{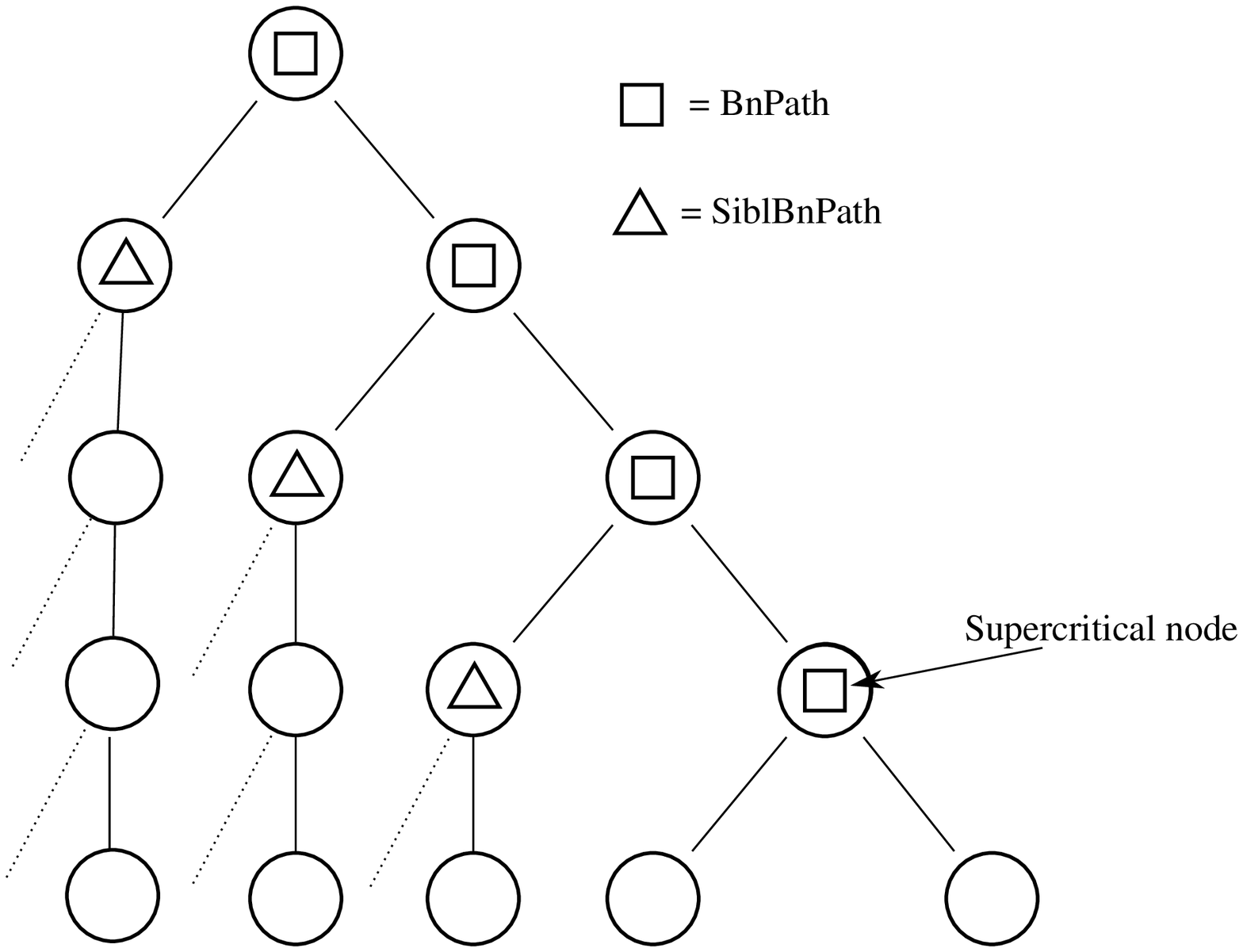}
\hspace{1cm} \vspace{.1cm}

Let $\preG_r$ be all the restrictions $g$ with domain $\Vars(\Ufixed^r)$. 
For every $g \in \preG_r$, for every internal node $i$ in $\Learnable_r$, we have that $v_{2i}^{g(E_r)}$ is defined since $g$ is defined for every $T_{2i}$ variable. 
For each $g \in \preG_r$ let $G_{r,g}$ be the set of extensions $g'$ of $g$ such that for all internal nodes $i$ in $\Learnable_r$,\, for all $a \not= v_{2i}^{g(E_r)}$ and all $b$,\, $g'$ is defined on $f_i(a,b)$, and $g'(E_r)$ is not empty. 
Finally, we take $\Ginit{r}$ to be $\bigcup_{g \in \preG_r} G_{r,g}$. The size of $\Ginit{r}$ is at most $k^{|\Vars|-M}$. 
 
Now fix $r \in R$ and $\ginit \in \Ginit{r}$ and let $D := \ginit(E_r)$. From this point on, we drop ``$r$'' from $\Learnable_r, \Learnable_r^*, \SiblBnPath_r, \BnPath_r$, and $\isc^r$. Since $r$ is thrifty for every $I$ in $D$, we have $v_{2\isc}^D\DEF$ and $v_{2\isc+1}^D\DEF$ (note $r$ queries the variable $f_{\isc}(v_{2\isc}^D,v_{2\isc+1}^D)$). 
Since we have now fixed $D = \ginit(E_r)$, when $g$ is an extension of $\ginit$ we just write $v_i^g$ and $X^g$ instead of $v_i^{g(D)}$ and $X^{g(D)}$. 

As in the proof of Theorem \ref{t:thrifty_advice}, we will define a procedure called \IntAdv\ (short for ``Interpret Advice'') that takes advice in the form of a $[k]$-string and interprets it as the code of an input in $D$. 
Ultimately we want to show: 
\def\footunique{Actually there is exactly one, but that is not important.}
\begin{prop} \label{p:IntAdvWorks}
For every $I \in D$, there is some restriction $g$ that extends $\ginit$ and some %\footnote{\footunique} 
advice $\adv^I$ of length at most $M-h$, such that $\IntAdv(\adv^I)\DEF = g$ and $I \in g(D)$ and $|\Dom(g) - \Dom(\ginit)| \ge |\adv^I| + h$.
\end{prop}
The procedure $\IntAdv$ is given precisely in pseudocode on page \pageref{pseudo:IntAdv} and relies on the subprocedures given on page \pageref{pseudo:Fill} and the following simple definition, which depends on the fixed input set $D$:
\begin{defn}[$g$ constrains $v_i$] \label{d:constrains}
We say $g$ constrains $v_i$ if for some $I \in g(D)$, the thrifty $i$ variable of $I$ is in $\Dom(g)$
\end{defn}

Recall how in the proof of Theorem \ref{t:thrifty_advice}, while reading the advice $\adv^I$ for $I \in D$, we maintain a current state $q \in \rPath[I]$ and build up a set of ``learned nodes'' which we called $\UL$. We are still building up a set of learned nodes, though in the pseudocode we have opted not to introduce a variable for that set explicitly. The learned nodes are just those nodes $j$ such that at some point during the execution of $\IntAdv(\adv^I)$,\, the subprocedure $\LearnNode$ is called with second argument $j$. In the thrifty proof, to characterize how we are interpreting the prefix of the advice that we have read so-far, we only need to record at most one value per node because every input is limited to querying its $n$ thrifty variables (in the pseudocode we used the variable $v^*$, a partial mapping from $T$ to $[k]$). More precisely, we had that if $v^*(i)\DEF = a$ after reading some advice elements $\vec{b}$, then $v_i^I = a$ for \emph{every} input $I$ in $E_r$ whose complete advice $\adv^I$ has $\vec{b}$ as a prefix, i.e. for every input in $v^*(E_r)$. Now that inputs can query non-thrifty variables, instead of $v^*$ we will be building up a restriction $g$, where initially $g = \ginit$. However, the meaning of $g(X)\DEF = a$ is what one would expect by analogy with $v^*$: if $g(X)\DEF = a$ after reading some advice elements $\vec{b}$, then $X^I = a$ for every input $I$ in $D$ whose complete advice $\adv^I$ has $\vec{b}$ as a prefix, i.e. for every input in $g(D)$. As with $v^*$ before, once we define the value $g$ takes on a given variable, we never change it. 

We first learn the children of $\isc$ at $r$; we treat this as a special case now because it is the only time when we learn two nodes while examining one state. After that we learn a node in essentially the same situation as before: we reach a state $q$ after reading some of the advice such that:
\begin{ppe} 
\item $q$ queries a variable $f_i(a_{2i},a_{2i+1})$ that is thrifty for every $I \in g(D)$, and \footnote{Here $g$ is the current restriction.} 
\item For $j =2i$ or $j=2i+1$ (not both), $g$ does not constrain $v_j$ ($j$ is the learned node).
\end{ppe}
We need $h-2$ such states after $r$ for each input in $D$. 
Let us say $q$ is a {\bf learning state} for $I \in D$ if both those conditions hold or if $q = r$.
In fact, by the properties of $\ginit$, and since after $r$ we will only ever learn nodes in $\Learnable^* = \bigcup_{j \in \SiblBnPath} \RightPath_j$, we can write the previous conditions in a more informative way: 
\begin{ppe}
\item For some internal $i \in \Learnable^* + (\BnPath - \isc)$,\ $q$ queries a variable $f_i(a_{2i},a_{2i+1})$ that is thrifty for every $I \in g(D)$, and 
\item If $i$ is in $\Learnable^*$ then $g$ does not constrain $v_{2i+1}$. \\
If $i$ is in $\BnPath - \isc$ and $j$ is the child of $i$ in $\BnPath$, then $g$ does not constrain $v_{\sibl(j)}$. 
\end{ppe}
We can be more specific still; later we will show that for each of the $h-2$ nodes $j \in \SiblBnPath$, we will learn at least one node in $\RightPath_j$. 

Let us now explain what ``learning a node'' entails. Temporarily fix $I \in D$. Suppose that while interpreting the advice for $I$ we reach a state $q \in \rPath^I$ that is a  learning state for $I$. So $q$ queries the variable $f_i(a_{2i},a_{2i+1})$ for some $i$ in $\Learnable^* + (\BnPath - \isc)$ and $a_{2i},a_{2i+1}$ in $[k]$. If $i$ is in $\BnPath - \isc$ then let $j$ be the child of $i$ in $\BnPath$, and otherwise let $j$ be $2i+1$. We are learning node $j$. 
If $j$ is an internal node, then first we use the advice, if necessary, to make $g$ total on $\Vars(T_{2j} + T_{2j+1})$. \label{line:secondFillPurposeExpl} After that, there is one variable $X$ that is the thrifty $j$ variable for every $I \in g(D)$. So then we ``learn'' $j$ by adding $X \mapsto a_j$ to $g$. The key point is that we have made progress since we used only $m = |\Dom(g) \ / \ \Vars(T_{2j} + T_{2j+1})|$ new elements of advice to define $g$ on $m+1$ new variables. 

\def\hq{\hat{q}}

The main thing we still need to show is that we can define $\adv^I$ so that $\IntAdv(\adv^I)$ will visit at least $h-2$ learning states for $I$ after $r$. 
As mentioned earlier, $I$ has at least one bottleneck node in $\RightPath_i$ for each of the $h-2$ nodes $i \in \SiblBnPath$. By Fact \ref{f:learnBnNodes}, for each of those bottleneck nodes $j$ there is a state $q^I_j$ in $\rPath[I]$ that queries the thrifty $\Par(j)$ variable of $I$, and no state between $r$ and $q^I_j$ that queries the thrifty $j$ variable of $I$. 

For each $i \in \SiblBnPath_r$, let $\hq_i^I$ be the earliest state in $\rPath[I]$ among the states 
\[\{ q^I_j \ | \ j \in \RightPath_i \text{ and $j$ is a bottleneck node of } I \}\]
and let $j_i$ be such that $\hq_i^I = q_{j_i}^I$. Then at least the nodes $\{j_i\}_{i \in \SiblBnPath}$ will be learned, and specifically $j_i$ will be learned upon reaching $q^I_{j_i}$. To prove this, for $\Par(j_i) \in \Learnable^*$ use Fact \ref{f:learnBnNodes} together with the comments given in footnote \ref{foot:bfoot} on page \pageref{foot:bfoot}. For $\Par(j_i) \in \BnPath - \isc$, use the following fact (with $j = \sibl(j_i)$ and $j' = \Par(j_i)$):
\begin{fact}\label{f:learnFromBnPath}
For all $I \in D$, if $j$ is a non-root node in $\BnPath$ and $j'$ is its ancestor in $\BnPath$, then there are states in $\rPath[I]$ that query the thrifty $j$ and $j'$ variables of $I$, and the first such state for $j$ occurs before the first such state for $j'$.
\end{fact}

\subsubsection*{Pseudocode for $\IntAdv$ and subprocedures}

The procedure $\Fill$ implements a very simple function: given inputs $g,V$ (the advice string $\s$ and the current index into it are implicit arguments), it just uses the advice to define $g$ on any variable in $V$ on which it is not yet defined. 
We call $\Fill$ in two qualitatively distinct situations. 
One is when for some $i$,\, $V$ is a single $i$ variable $X$ such that for every $I \in g(D)$, we have determined that either $i$ is not a bottleneck node of $I$ or $X$ is not thrifty for $I$. That is the situation when we call $\Fill$ from $\IntAdv$. The other situation occurs when $\LearnNode$ calls $\Fill$ on $\Vars(T_{2j} \cup T_{2j+1})$ for some $j$ that we have decided to learn. 
We do this because in order to learn $j$, we need $g$ to be defined on enough input variables that the inputs in $g(D)$ agree on the ``name'' of their thrifty $j$ variable, i.e. we need $v_{2j}^g\DEF$ and $v_{2j+1}^g\DEF$.

\vspace{10pt}

\noindent
{\bf Subprocedure} $\Fill(g \in G, V \subseteq \Vars)$:
\begin{algorithmic}[1] \label{pseudo:Fill}
  \STATE let $a_1,\ldots,a_m$ be the next $m = |V / \Dom(g)|$ elements of the advice string
  \STATE let $X_1,\ldots,X_{m}$ be $V / \Dom(g)$ sorted according to some globally fixed order on $\Vars$
  \STATE add $X_1 \mapsto a_1, \ldots, X_m \mapsto a_m$ to $g$
\end{algorithmic}

\vspace{7pt}

\vspace{3pt}
\noindent
{\bf Subprocedure} $\LearnNode( g \in G, j \in \Learnable^*, b \in [k])$:
\begin{algorithmic}[1] \label{pseudo:LearnNode}
  \IF{$j$ is not a leaf}
     \STATE $\Fill(g,\Vars(T_{2j} + T_{2j+1}))$ \label{line:makeTotalSubtrees}
     \STATE let $X := f_j(v_{2j}^{g},v_{2j+1}^{g})$
  \ELSE 
     \STATE let $X = l_j$
  \ENDIF
  \STATE add $X \mapsto b$ to $g$ \label{line:addToG}
\end{algorithmic}

\vspace{10pt}

\def\afoot{This makes sense because every node in $\BnPath$ other than $\isc$ has a child in $\BnPath$.}

\def\bfoot{$v_{2i}^{g}\DEF$ by definition of $\ginit$ since $2i$ is the left child of a node in $\Learnable$. Also $X = f_i(v_{2i}^g,b)$ for some $b$ since $\ginit$ is not defined on $X$. Also $v_{2i+1}^{g}\DEF = b$ -- since $X$ is not wrong-wrong for any $I \in g(D)$, it must be thrifty for every $I \in g(D)$.}

\def\gets{\leftarrow}

\noindent 
{\bf Procedure $\IntAdv(\s \in [k]^*)$: } 
\begin{algorithmic}[1] \label{pseudo:IntAdv}
  \STATE \COMMENT{Note the advice string $\s$ and the current index into it are implicit arguments in every call to $\Fill$ and $\LearnNode$.}
  \STATE $q \gets r,\ g \gets \ginit$      
  \WHILE{ $q$ is not an output state }
     \STATE $i \gets \node(q)$,\ $X \gets \var(q)$
     \IF{$X \not \in \Dom(g)$} \label{line:mainCond}
       \IF{$i = \isc$} \label{line:beginMainBlock}
	   \STATE add $l_{2\isc} \mapsto v_{2\isc}^{\ginit}$ and $l_{2\isc+1} \mapsto v_{2\isc+1}^{\ginit}$ to $g$ \label{line:firstLearn}
       \ELSIF{$i \in \BnPath - \isc$ }  
	   \STATE let $j$ be the child of $i$ in $\BnPath$ \label{lin:needs_just_a} \footnote{ \afoot } 
           \STATE let $a_{2i},a_{2i+1}$ be such that $X = f_i(a_{2i},a_{2i+1})$
           \IF{$v_j^{g}\DEF = a_j$ and $g$ does not constrain $v_{\sibl(j)}$}
	      \STATE \COMMENT{Uses $|\Vars({\sf descendants}(\sibl(j))) \ / \ \Dom(g)|$ elements of advice:}
              \STATE $\LearnNode(g,\sibl(j),a_{\sibl(j)})$ \label{line:learnbn}
           \ELSE
              \STATE $\Fill(g,\{X\})$ \COMMENT{Uses one element of advice.}
           \ENDIF
       \ELSE[$i \in \Learnable^*$] \label{line:notBnPathBlockStart} 
  	   \IF{ $i$ is an internal node and $g$ does not constraint $v_{2i+1}$ } 
             \STATE let $b$ be such that $X = f_i(v_{2i}^{\ginit}, b)$ \footnote{ \bfoot \label{foot:bfoot}}
             \STATE \COMMENT{Uses $|\Vars({\sf descendants}(2i+1)) \ / \ \Dom(g)|$ elements of advice:}
	     \STATE $\LearnNode(g,2i+1,b)$  \label{line:learnother}
          \ELSE 
             \STATE $\Fill(g,\{X\})$  \COMMENT{Uses one element of advice.}
          \ENDIF
        \ENDIF \label{line:notBnPathBlockEnd} \label{line:endMainBlock}
      \ENDIF 
  \STATE $q \gets $  the state reached by taking the edge out of $q$ labeled $g(X)$
  \ENDWHILE
  \STATE return $g$
\end{algorithmic} 
\end{proof}

%% file: general_less_thrifty.tex
 The previous result can be generalized to give gradually weaker lower bounds for gradually weaker restrictions on the model.
 For $B$ a deterministic BP that solves $\BT$, for every state $q$ of $B$ that queries a variable $f_i(a,b)$, let $\RightThrifty(q)$ be the set of integers $a'$ (including $a$) such that there is some input to $B$ that visits
 $q$ and has values $a'$ and $b$ for nodes $2i$ and $2i+1$. Likewise, let $\LeftThrifty(q)$ be the set of integers $b'$ such that there is some input that visits $q$ and has values $a$ and $b'$ for nodes $2i$ and $2i+1$. Theorem \ref{t:less_thrifty} is the special case of the following result when $\pi = 1$.
 \begin{theorem} \label{t:param_less_thrifty}
 For any $h,k \ge 2$ and $\pi < k$, if $B$ is a deterministic BP that solves $\BT$ such that $|\LeftThrifty(q)| \le \pi$ and $|\RightThrifty(q)| \le \pi$ for every state $q$ that queries an internal node, then $B$ has at least $k^h / \pi^{h-2}$ states.
 \end{theorem}
 
 \begin{proof}
  We modify the proof of Theorem \ref{t:less_thrifty}. We first need to verify that the analogue of Lemma \ref{l:basic_less_thrifty} for this context holds:
 \begin{lemma}\label{l:basic_param_less_thrifty}
   For any $I$ and internal node $i$, there is at least one state $q$ on the computation path of $I$ that queries the thrifty $i$ variable of $J$, and for every such $q$, for each child $j$ of $i$, there is a state on the computation path of $I$ before $q$ that queries the thrifty $j$ variable of $I$.
 \end{lemma}
 \begin{proof}
  We use the strategy from the proof of Lemma \ref{l:basic_thrifty} on page \pageref{l:basic_thrifty}. $I$ must visit at least one state that queries its thrifty root variable, since otherwise $B$ would make a mistake on an input $J$ that is identical to $I$ except $f_1^{J}(v_2^I,v_3^I) = k - f_1^I(v_2^I,v_3^I)$. Now let $q$ be a state on the computation path of $I$ that queries the thrifty $i$ variable of $I$, for some internal node $i$. Suppose the lemma does not hold for this $q$, and wlog assume there is no earlier state that queries the thrifty $2i$ variable of $I$. For every $a \not = v_{2i}^I$ there is an input $J_a$ that is identical to $J$ except $v_{2i}^{J_a} = a$. This implies $|\RightThrifty(q)| = k$, contradicting the assumption that  $|\RightThrifty(q)| \le \pi < k$.
 \end{proof}
 The assignment of pebbling sequences to inputs and the definition of supercritical states is the same. 
In fact nothing more needs to be changed until the statement of the Main Lemma, which is now:
 \begin{lemma}[Main Lemma]\label{l:param_less_thrifty_main_one}
  For some integer $M$, for every supercritical state $r \in R$, there is a set of restrictions $\Ginit{r}$ of size at most $k^{|\Vars| - M}$ such that $\{\ginit(E_r) \}_{\ginit \in \Ginit{r}}$ is a partition of $E_r$ and for every $\ginit$ in $\Ginit{r}$, there is an injective function from $\ginit(E_r)$ to $[\pi]^{h-2} \times [k]^{M-h}$.
 \end{lemma}
 So in order to cope with the relaxed restrictions on the model, in addition to the $[k]$-valued advice string of length $M-h$ we now have a $[\pi]$-valued advice string of length $h-2$. One can show the theorem follows from the lemma in the same way as in the proof of Theorem \ref{t:less_thrifty}. Really at this point there is just one additional observation needed to adapt the proof of Theorem \ref{t:less_thrifty}: 
Suppose we have a set of inputs $F$ all of which have value $a$ for $v_{2i}$ (i.e. $v_{2i}^{F}\DEF = a$), and all the inputs in $F$ visit a state $q$ that queries a variable $f_i(a,b)$. Then we can use the elements of $[\pi]$ to code the values of $v_{2i+1}$ for inputs in $F$. More concretely, let $a_1,\ldots,a_m$ be the $m \le \pi$ integers $\LeftThrifty(q)$ in increasing order. Then to each $I \in F$ we assign the index of $v_{2i+1}^I$ in $a_1,\ldots,a_m$. Of course a similar property holds for the case when $F$ is a set of inputs that agree on $v_{2i+1}$. We use this observation later to show that if we ``know'' the value of node $2i$ upon reaching $q$, then we can learn node $2i+1$ with the help of just an element of $\pi$-valued advice, and similarly for learning node $2i$. 
 
 The definition of $\Ginit{r}$ is the same, and as before we fix $r \in R$ and $\ginit \in \Ginit{r}$ 
 and then define a procedure that interprets some given advice as the code of an input in $D := \ginit(E_r)$. The analogue of Proposition \ref{p:IntAdvWorks} (page \pageref{p:IntAdvWorks}) is:
 \begin{prop} \label{p:paramIntAdvWorks}
 For every $I \in D$, there is some restriction $g$ that extends $\ginit$, a $[\pi]$-valued advice string $\advpi^I$ of length $h-2$ and a $[k]$-valued advice string $\advk^I$ of length at most $M-h$, such that $\IntAdv(\advk^I, \advpi^I)\DEF = g$ and $I \in g(D)$ and $|\Dom(g) - \Dom(\ginit)| \ge |\adv^I| + h$.
 \end{prop}

However, it will be convenient to instead give a procedure $\IntAdv'$ for which the following superficially different proposition holds:
 \begin{prop} \label{p:altParamIntAdvWorks}
 For every $I \in D$, there is some restriction $g$ that extends $\ginit$, a $[\pi]$-valued advice string $\advpi^I$ of length at least $h-2$ and 
 a $[k]$-valued advice string $\advk^I$ of length at most $M-|\advpi^I| - 2$, such that $\IntAdv'(\advk^I, \advpi^I)\DEF = g$ and $I \in g(D)$ and $|\Dom(g) - \Dom(\ginit)| \ge |\adv^I| + |\advpi^I| + 2$.
 \end{prop}
 To get the procedure $\IntAdv$ of Proposition \ref{p:paramIntAdvWorks} from the procedure $\IntAdv'$ of Proposition \ref{p:altParamIntAdvWorks}, just run $\IntAdv'$ until $h-2$ elements of the $[\pi]$-valued advice have been used, and then, if necessary, use elements of the $[k]$-valued advice whenever an additional element of the $[\pi]$-valued advice is required. This works since $\pi \le k$ and $|\advk^I| \le M-|\advpi^I| - 2$.
 
 Let us say {\bf $q$ is right-thrifty for $I$} if $q$ queries a variable $f_i(a,b)$ such that $b = v_{2i+1}^I$ and $a \not = v_{2i}^I$. Similarly define {\bf left-thrifty for $I$}. Previously, while interpreting the advice for $I$ we only learned node values at states that are thrifty for $I$. Now we may learn node values at states that are thrifty, right-thrifty, or left-thrifty for $I$. As before, we always learn the children of $\isc$, and the remaining $h-2$ nodes we learn are in $\Learnable^*$. 
 
 First we consider the case of learning a node in $\SiblBnPath$. We consider the case of learning a left child $2i$ -- the case of learning a right child is similar. Let $q$ be the first state in $\rPath[I]$ that queries the thrifty $2i+1$ variable of $I$. \emph{If} we learn $2i$, then we do so at the first state $q'$ after $q$ that queries an $i$ variable that is thrifty or right-thrifty for $I$.
 Now we consider the case of learning a node in $\Learnable^* - \SiblBnPath$. Every node in $\Learnable^* - \SiblBnPath$ is a right child, so suppose we are learning $2i+1$. Then we do so at the first state in $\rPath[I]$ that queries an $i$ variable that is thrifty or left-thrifty for $I$.
 
 As before, for each $I$ in $D$ and each of the $h-2$ nodes $i$ in $\SiblBnPath$, we will learn at least one node in $\RightPath_i$ (and of course we still learn the children of the supercritical node $\isc$). This is again proved using Facts \ref{f:learnBnNodes} (page \pageref{f:learnBnNodes}) and \ref{f:learnFromBnPath} (page \pageref{f:learnFromBnPath}); both still hold since we did not change the assignment of pebbling sequences to inputs.

 We provide pseudocode for $\IntAdv'$, just in case the reader has questions not explicitly addressed in the preceding prose.  
 On the other hand, there is little to read since it {\bf differs from the previous definition of $\IntAdv$ (\ref{pseudo:IntAdv} on page \pageref{pseudo:IntAdv}) only in a few lines} near the two calls to $\LearnNode$ (specifically lines \ref{line:startdiffa} - \ref{line:enddiffa} and \ref{line:startdiffb} - \ref{line:enddiffb}). The two subprocedures $\Fill$ and $\LearnNode$ do not use the $[\pi]$-valued advice and do not need to be modified. 

\vspace{10pt}

\noindent 
{\bf Procedure $\IntAdv'(\sk \in [k]^*, \spi \in [\pi]^*)$: } 
\begin{algorithmic}[1] \label{pseudo:ParamIntAdv}
  \STATE \COMMENT{Note the advice string $\sk$ and the current index into it are both implicit arguments in every call to $\Fill$ and $\LearnNode$.}
  \STATE $q \gets r,\ g \gets \ginit$    
  \WHILE{ $q$ is not an output state }
     \STATE $i \gets \node(q)$,\ $X \gets \var(q)$
     \IF{$X \not \in \Dom(g)$}
       \IF{$i = \isc$} 
	   \STATE add $l_{2\isc} \mapsto v_{2\isc}^{\ginit}$ and $l_{2\isc+1} \mapsto v_{2\isc+1}^{\ginit}$ to $g$
       \ELSIF{$i \in \BnPath - \isc$ }  
	   \STATE let $j$ be the child of $i$ in $\BnPath$ 
           \STATE let $a_{2i},a_{2i+1}$ be such that $X = f_i(a_{2i},a_{2i+1})$
           \IF{$v_j^{g}\DEF = a_j$ and $g$ does not constrain $v_{\sibl(j)}$}
              \STATE let $z$ be the next element of the $[\pi]$-valued advice. \label{line:startdiffa}
              \STATE if $j = 2i+1$ then let $b$ be the $z$-th greatest integer in $\RightThrifty(q)$ and otherwise let $b$ be the $z$-th greatest integer in $\LeftThrifty(q)$
	       \STATE \COMMENT{Uses $|\Vars({\sf descendants}(\sibl(j))) \ / \ \Dom(g)|$ elements of $[k]$-valued advice:}
              \STATE $\LearnNode(g,\sibl(j),b)$  \label{line:enddiffa}
           \ELSE
              \STATE $\Fill(g,\{X\})$ \COMMENT{Uses one element of $[k]$-valued advice.}
           \ENDIF
       \ELSE[$i \in \Learnable^*$] 
  	   \IF{ $i$ is an internal node and $g$ does not constraint $v_{2i+1}$ } 
	     \STATE let $z$ be the next element of the $[\pi]$-valued advice.  \label{line:startdiffb}
             \STATE let $b$ be the $z$-th greatest integer in $\LeftThrifty(q)$.
             \STATE \COMMENT{Uses $|\Vars({\sf descendants}(2i+1)) \ / \ \Dom(g)|$ elements of $[k]$-valued advice:}
	     \STATE $\LearnNode(g,2i+1,b)$  \label{line:enddiffb}
          \ELSE 
             \STATE $\Fill(g,\{X\})$  \COMMENT{Uses one element of $[k]$-valued advice.}
          \ENDIF
        \ENDIF 
      \ENDIF 
  \STATE $q \gets $  the state reached by taking the edge out of $q$ labeled $g(X)$
  \ENDWHILE
  \STATE return $g$
\end{algorithmic} 

\end{proof}

We give one more extension of the thrifty lower bound. 
We introduce another parameter $w$: for each input $I$, we require that there are at most $w$ nodes $i$ such that $I$ visits a state $q$ with $|\RightThrifty(q)| > 1$ or $|\LeftThrifty(q)| > 1$.
The motivation for this is that for $w = 1$ and $\pi = \log k - \log \log k$, the model includes BPs that achieve the best known upper bounds for $\BT$, namely $O( k^h / \log k )$. For those parameters the theorem gives a lower bound of $k^h / (\log k - \log \log k) = \Omega(k^h / \log k)$. In \cite{mfcs} it was shown that the minimum number of states for \emph{unrestricted} deterministic BPs solving $\BT[3][k]$ is $\Theta(k^3 / \log k)$.

\begin{theorem} \label{t:two_param_less_thrifty}
 For any $h,k \ge 2$ and $\pi < k$ and $w < h-2$, if $B$ is a deterministic BP that solves $\BT$ such that $|\LeftThrifty(q)| \le \pi$ and $|\RightThrifty(q)| \le \pi$ for every state $q$ that queries an internal node, and such that for every input $I$ there are at most $w$ nodes $i$ such that $I$ visits a state $q$ that queries an $i$-variable and has $|\RightThrifty(q)| > 1$ or $|\LeftThrifty(q)| > 1$, then $B$ has at least $k^h / \pi^{w}$ states.
 \end{theorem}
\begin{proof}
 This is an easy modification of the proof of the previous result. Note that $\RightThrifty(q)$ and $\LeftThrifty(q)$ are properties of $B$, so independently of the advice we can label each internal node querying state $q$ with the quanities $|\RightThrifty(q)|$ and $|\LeftThrifty(q)|$. If for some advice $q$ is a learning state, then we will use an element of the $\pi$-valued advice iff $|\RightThrifty(q) > 1|$ or $|\LeftThrifty(q) > 1|$. Hence, for any input $I$ in $E_r$, we can define the advice for $I$ so that for all but at most $w$ of the $h-2$ learning states $q$ of $I$ after $r$, we do not need to use an element of the $[\pi]$-valued advice to learn a child of $\node(q)$. So we only need a $[\pi]$-valued advice string of length $w$.
\end{proof}

%% file: masters_paper.bbl
\newcommand{\etalchar}[1]{$^{#1}$}
\begin{thebibliography}{BCM{\etalchar{+}}09b}

\bibitem[BCM{\etalchar{+}}09a]{fsttcs}
Mark Braverman, Stephen Cook, Pierre McKenzie, Rahul Santhanam, and Dustin
  Wehr.
\newblock Fractional pebbling and thrifty branching programs.
\newblock In Ravi Kannan and K~Narayan Kumar, editors, {\em IARCS Annual
  Conference on Foundations of Software Technology and Theoretical Computer
  Science (FSTTCS 2009)}, volume~4 of {\em Leibniz International Proceedings in
  Informatics (LIPIcs)}, pages 109--120, Dagstuhl, Germany, 2009. Schloss
  Dagstuhl--Leibniz-Zentrum fuer Informatik.

\bibitem[BCM{\etalchar{+}}09b]{mfcs}
Mark Braverman, Stephen~A. Cook, Pierre McKenzie, Rahul Santhanam, and Dustin
  Wehr.
\newblock Branching programs for tree evaluation.
\newblock In Rastislav Královic and Damian Niwinski, editors, {\em MFCS},
  volume 5734 of {\em Lecture Notes in Computer Science}, pages 175--186.
  Springer, 2009.

\bibitem[BCM{\etalchar{+}}09c]{manuscript}
Mark Braverman, Stephen~A. Cook, Pierre McKenzie, Rahul Santhanam, and Dustin
  Wehr.
\newblock Pebbles and branching programs for tree evaluation.
\newblock A draft manuscript, available on line at
  {\verb+http://www.cs.toronto.edu/~sacook/homepage/pebbles.pdf+}, 2009.

\bibitem[Coo74]{co74}
S.~Cook.
\newblock An observation on time-storage trade off.
\newblock {\em J. Comput. Syst. Sci.}, 9(3):308--316, 1974.

\bibitem[CS76]{cose76}
S.~Cook and R.~Sethi.
\newblock Storage requirements for deterministic polynomial time recognizable
  languages.
\newblock {\em J. Comput. Syst. Sci.}, 13(1):25--37, 1976.

\bibitem[Kla85]{klawe}
M.~Klawe.
\newblock A tight bound for black and white pebbles on the pyramid.
\newblock {\em J. ACM}, 32(1):218--228, 1985.

\bibitem[Ne{\u{c}}66]{ne66}
{\`{E}}.~Ne{\u{c}}iporuk.
\newblock On a boolean function.
\newblock {\em Doklady of the Academy of the USSR}, 169(4):765--766, 1966.
\newblock English translation in \emph{Soviet Mathematics Doklady} 7:4, pp.\
  999-1000.

\bibitem[Nor09]{nordstrom}
J.~Nordstr\"{o}m.
\newblock New wine into old wineskins: A survey of some pebbling classics with
  supplemental results.
\newblock Available on line at
  {\verb+http://people.csail.mit.edu/jakobn/research/+}, 2009.

\bibitem[PH70]{pahe70}
M.~Paterson and C.~Hewitt.
\newblock Comparative schematology.
\newblock In {\em Record of Project MAC Conference on Concurrent Systems and
  Parallel Computations}, pages 119--128, 1970.
\newblock (June 1970) ACM. New Jersey.

\bibitem[Raz91]{ra91}
A.~Razborov.
\newblock Lower bounds for deterministic and nondeterministic branching
  programs.
\newblock In {\em 8th Internat. Symp. on Fundamentals of Computation Theory},
  pages 47--60, 1991.

\end{thebibliography}
